\documentclass[aps,prx,superscriptaddress,twocolumn]{revtex4-2}
 \usepackage{graphicx,color}
 \usepackage{verbatim}
 \usepackage{amssymb}   % for math
 \usepackage{amsmath}
 \usepackage{amsfonts}
 \usepackage{mathdots}
 \usepackage{hyperref}
 \usepackage{epsfig}
 \usepackage{hyperref}
 \usepackage{xcolor}
 \usepackage{cancel}
 \usepackage{booktabs}
 \usepackage{array}
 \usepackage{array, makecell} 
 \definecolor{myblue}{RGB}{46, 48,146}
 \usepackage{multirow}
\hypersetup{colorlinks=true,linkcolor=myblue,citecolor=myblue,urlcolor=myblue, linktocpage}
 \usepackage{braket}
 \usepackage{bm}
 \usepackage{multirow}
 
\begin{document}
 	\title{Phase shifts, band geometry and responses in triple-Q charge and spin density waves}
 	
 	\author{Ying-Ming Xie}\thanks{yingming.xie@riken.jp}
    \affiliation{RIKEN Center for Emergent Matter Science (CEMS), Wako, Saitama 351-0198, Japan} 	
    \author{Naoto Nagaosa}\thanks{nagaosa@riken.jp}
   \affiliation{RIKEN Center for Emergent Matter Science (CEMS), Wako, Saitama 351-0198, Japan} 
 \affiliation{Fundamental Quantum Science Program, TRIP Headquarters, RIKEN, Wako 351-0198, Japan}
\begin{abstract}
%Recently, there has been growing interest in the impacts of phase shifts within the triple-Q spin density wave (SDW) order parameters. Concurrently, it is widely recognized that incommensurate triple-Q charge density waves (CDW) are also prevalent in low-dimensional materials, where the phase degrees of freedom in the order parameters are generally allowed.  In this study, we systematically investigate the pivotal effects arising from both triple-Q CDW and SDW order parameters, with particular consideration given to possible phase shifts.  We show that the phase shifts play a crucial role in determining the real-space topology of triple-Q density waves. More importantly,  we show that the triple-Q CDW and SDW order parameters would influence the band geometry in the momentum space,  where multiband Dirac-like fermions are induced by the triple-Q density wave order parameters near the Fermi energy. Furthermore, we explicitly establish that such nontrivial band geometry, combined with symmetry-breaking induced by phase shifts, leads to a variety of intriguing linear and nonlinear responses. 
Triple-Q density waves are commonly found in various materials, such as charge density waves in transition metal dichalcogenides and spin density waves (skyrmion crystals) in B20 compounds. Compared to single-Q density waves, triple-Q density waves possess an additional internal degree of freedom—a phase shift arising from the phase of the order parameters, in addition to the translations of the density waves. In this study, we systematically investigate the significant effects stemming from both triple-Q CDW and SDW order parameters, with particular emphasis on potential phase shifts. We demonstrate that these phase shifts play a crucial role in influencing the interference effects of triple-Q density waves on electronic states. Due to such interference, the band geometry in the momentum space becomes nontrivial at the hot spots,  where multiband Dirac-like fermions are induced near the Fermi energy. Furthermore, we explicitly establish that the nontrivial band geometry, combined with symmetry-breaking induced by phase shifts, leads to a variety of intriguing linear and nonlinear responses.

 	\end{abstract}
 	
 	\date{\today}
 	
 	\maketitle

%The universal scaling behavior is an important concept in statistical physics, where various models can be classified in terms of their collective behavior.  Scaling phenomena are also fundamental in condensed matter physics, especially in studying critical phase transitions, where their lareg-scale behavior is insensitive to irrelevant microscopic details. The universal scaling behavior can also exist in studying the statics and dynamics of elastic systems in a random environment \cite{Fisher1993, Pierre}. These are two type for such systems: (i) non-periodic type, such as magnetic domain walls \cite{Lemerle1998, Kim2009}, invasion in porous media \cite{Wilkinson1983, Rost2007}, or epitaxial growth \cite{barabási_stanley_1995}; (ii) periodic type, such as superconducting votex lattices \cite{Schmid1973TY, Blatter1994, Thomann2012}, charge density waves \cite{Sneddon1982, Fisher1985,Narayan1992,Narayan1992PRL,Threshold1993,Gruner1988}.  The universal scaling behaviours in these elastic systems have attracted wide interest.
\emph{Introduction---} For the incommensurate density waves,  there are possible phase degrees of freedom. For example, the incommensurate density waves exhibit an exotic gapless collective excitation --- phason. These phasons can be excited with an electric field beyond the threshold,  leading to the so-called depinning phenomena. Such phenomena have been observed in both charge density wave (CDW) and spin density wave (SDW) systems \cite{LEE1974,Patrick_Lee1,Patrick_Lee2,Sneddon1982, Gruner1988, Gruner1994,Reichhardt2022}.  Moreover, the CDWs and SDWs often share similar physics with each other.  Recently, considerable efforts have been devoted to identifying the common physics shared by various density wave systems \cite{Yingming2023, Yingming2024, Birch}.

Unlike single-Q density waves, triple-Q density waves further exhibit an internal phase shift degree of freedom made from the phase of the order parameters, which determines the structure of triple-Q density waves. The phase shifts are naturally expected to appear in multi-Q incommensurate density waves as well. Especially, the incommensurate triple-Q CDW systems are widely recognized in low-dimensional materials, such as layered transition-metal dichalcogenides (TMDC) 1T TaS$_2$, 1T TiSe$_2$, 2H TaSe$_2$ \cite{McMillan1976, Rossnagel_2011}, while triple-Q SDWs are commonly found in noncolinear magnetic materials, such as skymrion crystals (SkX) \cite{Boni2009,Yu2010,Nagaosa2013,Tokura2021}. Moreover, the important effects of phase shifts on the SkX were recently highlighted \cite{Takashi2019, Hayami2021,Shimizu2022}, which play a crucial role in affecting the scalar spin chirality and lowering the rotational symmetry in SkX \cite{Hayami2021}. However, theoretical work that studies the impacts of the interference of triple-Q density waves with phase shifts on the electronic states in a unified fashion is still lacking. %a theoretical work that simultaneously describes the effects of phase shifts in both incommensurate triple-Q CDW and SDW systems is still lacking.

%Moreover, the recent experimental and theoretical progress \cite{Tomohiro2024, Senthil2023} also motivates the study of nontrivial band geometry in density wave systems.

In this work, we present a minimal theory to capture the phase shifts, band geometry, and their interplay in various linear and nonlinear responses in two-dimensional triple-Q density wave systems. The theoretical framework is applied equally to both triple-Q CDWs and SDWs. We demonstrate that the interference of triple-Q density waves can influence the electronic band geometry at the hot spots, creating multiband Dirac Hamiltonians. Especially, we highlight that phase shifts play a critical role in modifying the geometry of low-energy bands.  We also explicitly show that the nontrivial band geometry, along with possible inversion or rotational symmetry breaking induced by phase shifts, enables a rich variety of linear and nonlinear responses in incommensurate triple-Q CDW and SDW systems.  

\emph{Symmetry classifications of triple-Q density waves---}
Let us first present the form of triple-Q density waves in two dimensions from the symmetry point of view. Without loss of generality, the three Q vectors are defined as $\bm{Q}_1=(1,0)Q, \bm{Q}_2=(-\frac{1}{2},\frac{\sqrt{3}}{2})Q, \bm{Q}_3=(-\frac{1}{2},-\frac{\sqrt{3}}{2})Q$, where the angle with respect to each other are $120$  $^{\circ}$.  The phase shifts along $\bm{Q}_{\nu}$  direction are defined as $\theta_{\nu}$. The structure of CDW/SDW is then determined by $\Theta=\frac{1}{\sqrt{3}}(\theta_1+\theta_2+\theta_3)$, while the other two degrees of freedom given by $\theta_X=\frac{1}{\sqrt{6}}(2\theta_1-\theta_2-\theta_3)$,  $\theta_Y=\frac{1}{\sqrt{2}}(\theta_2-\theta_3)$. Here, $\theta_{X(Y)}$ 
represent the phasons, which are typical low-energy elementary excitations in density wave systems upon translational motion \cite{Hoshino2018,Fukuyama2014}. In this work, to preserve $C_{3z}$ symmetry, we would set  $\theta_1=\theta_2=\theta_3$. But the coupling to the two phasons given by $\theta_X$ and $\theta_Y$ is an interesting problem for future study.  As shown in Table I, we can classify triple-Q density waves using the irreducible representations of the $C_3$ point group, where the spin Pauli matrices are defined as $\sigma_j$, the phase shift is defined as $\theta$, and the position vector is $\bm{r}$.   
 \begin{table}
 \caption{Classifications of triple-Q density waves using the irreducible representations of the $C_{3z}$ point group. Here, $\omega=e^{i2\pi/3}$,  $\bm{Q}_1=(1,0)Q, \bm{Q}_2=(-\frac{1}{2},\frac{\sqrt{3}}{2})Q, \bm{Q}_3=(-\frac{1}{2},-\frac{\sqrt{3}}{2})Q$, $\sigma_j$ are  spin  Pauli matrices. }
  \begin{ruledtabular}
 	\begin{tabular}{|c|c|c|}
 		$C_3$ & 	basis functions & spin matrices\\\hline
 		1& $f_{A}(\bm{r})=	\sum_{\nu=1}^3 \cos(\bm{Q}_{\nu}\cdot \bm{r}+\theta)$& $\sigma_0, \sigma_z$ \\\hline
 		$\omega$&  $f_{E_1}(\bm{r})=\sum_{\nu=1}^3 \omega^{\nu-1}\sin(\bm{Q}_{\nu}\cdot \bm{r}+\theta)$&$\sigma_{+}=\sigma_x+i\sigma_y$\\\hline
 		$\omega^*$& $f_{E_2}(\bm{r})=\sum_{\nu=1}^3 \omega^{1-\nu}\sin(\bm{Q}_{\nu}\cdot \bm{r}+\theta)$&$\sigma_{-}=\sigma_x-i\sigma_y$\\
 	\end{tabular}
  \end{ruledtabular}
 \end{table}
 In the spinless case, the $C_{3z}$ invariant triple-Q density wave order parameter, i.e., CDW order parameter, is 
 \begin{equation}
 	\Delta_{\text{CDW}}(\bm{r})=2\Delta f_{A}(\bm{r})=2\Delta\sum_{\nu=1}^3\cos(\bm{Q}_{\nu}\cdot \bm{r}+\theta_{\nu}).
 \end{equation}
 where $\Delta$ denotes the CDW potential, and $f_A(\bm{r})$ is the basis function of $A$ irreducible representation of $C_3$ point group.  In the spinful case, the $C_{3z}$ invariant triple-Q density wave order parameter, i.e., SDW order parameter, is obtained as  
 \begin{equation}
     \Delta_{\text{SDW}}(\bm{r})=\frac{\Delta}{2}[f_{E_1}(\bm{r})\sigma_{-}+f_{E_2}(\bm{r})\sigma_{+}]-\Delta f_{A}(\bm{r})\sigma_z. \label{SDW}
 \end{equation}
Here, $f_{E_1}(\bm{r})$ and $f_{E_2}(\bm{r})$ (see Table I) are the basis functions of $E_1$ and $E_2$ irreducible representations of $C_3$ group.  We can rewrite the form of $\Delta_{\text{SDW}}(\bm{r})$ as
 \begin{equation}
 	\Delta_{\text{SDW}}(\bm{r})=\sum_{i} \bm{S}_i^{\text{spiral}}\cdot \bm{\sigma},
 \end{equation}
and obtain the triple-Q spiral spin textures \cite{Hayami2021}
 \begin{equation}
 	\bm{S}_i^{\text{spiral}}=\Delta \sum_{\nu=1}^{3} (\sin \mathcal{Q}_{\nu} \cos \phi_{\nu}, \sin \mathcal{Q}_{\nu} \sin \phi_{\nu}, -\cos \mathcal{Q}_{\nu})
 \end{equation}
 with $\mathcal{Q}_{\nu}=\bm{Q}_{\nu}\cdot \bm{r}_i+\theta_\nu$, $\phi_{\nu}=\frac{2\pi}{3}(\nu-1)$. Here, $\bm{Q}_{\nu}$ is the  wave vector of $\nu$-th spiral, and $\theta_{\nu}=\theta$. For the sake of simplicity, we neglect the spin-orbit interaction in our model so that the direction of the spin-rotating plane does not matter.    We shall also focus on the simplest case that the order parameter only exhibits the spatial and spin degree of freedom.  %If more degrees of freedom are involved in the density wave order parameters, such as sublattice, other exotic CDWs and SDWs may be also allowed. 
 %It is worth noting that the same form of spiral spin textures was also introduced in studying the skyrmion crystals [].

\begin{figure}
	\centering
	\includegraphics[width=1\linewidth]{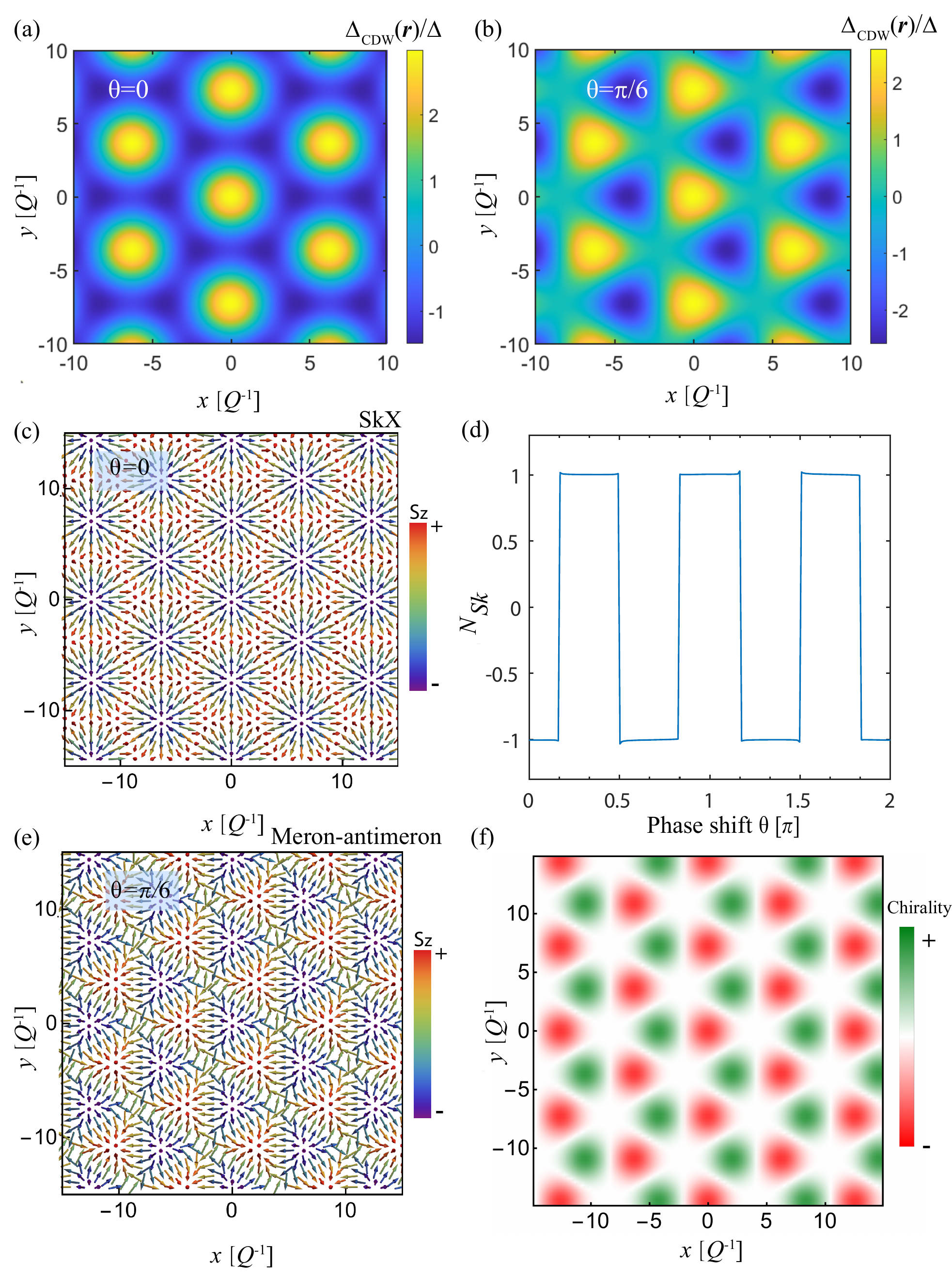}
	\caption{ (a) and (b) The real-space dependence of triple-Q CDW potential $\Delta_{\text{CDW}}(\bm{r})$ at $\theta=0$ and $\theta=\pi/6$, respectively. (c) and (e) The triple-Q spiral spin textures at $\theta=0$ and $\theta=\frac{\pi}{6}$, respectively.  (d) The total skyrmion charge $N_{sk}$ of each unit cell versus the phase shift $\theta$ calculated from the normalized spin vectors $\bm{s}$.  (f) The real-space spin chirality $\chi=\bm{S}_i\cdot(\bm{S}_j\times \bm{S}_k)$, where the site indcies $i,j,k$ are in counterclockwise order. }
	\label{fig:fig1}
\end{figure}

\emph{Real space landscape and topology of triple-Q density waves---}
To visualize these order parameters and highlight the effects of the phase shift $\theta$ on the triple-Q density waves, we plot the real space landscape of triple-Q CDW and triple-Q SDW with various $\theta$ in Fig.~\ref{fig:fig1}. 

Figs.~\ref{fig:fig1}(a) and (b) display the CDW order parameter $\Delta_{\text{CDW}}(\bm{r})$ at $\theta=0$ and $\theta=\pi/6$. Interestingly, it can be seen that the presence of finite $\theta$ can break the inversion symmetry.  
To show the real space feature of SDW order $\Delta_{\text{SDW}}(\bm{r})$, the spiral spin textures given by $\bm{S}^{\text{spiral}}$ at $\theta=0$ is plotted in  Fig.~\ref{fig:fig1}(c), which displays as a skyrmion crystal. Then we study the topology of the spiral spin textures more explicitly. The calculated skyrmion charge $N_{sk}=\frac{1}{4\pi}\int\int d^2\bm{r} \bm{s}\cdot(\partial_x \bm{s} \times \partial_y \bm{s} )$ as a function of the phase shift $\theta$ with $\bm{s}$ being the  unit vector of spin normalized from $\bm{S}$ is shown in  Fig.~\ref{fig:fig1}(d). Interestingly, we find that the skyrmion charge $N_{sk}$ is quantized at $\pm 1$ except for  flipping sign at $\theta=\frac{(2n+1)\pi}{6}$ ($n$ are integers). At these angles where $N_{sk}$ changes sign, it can be seen that the spiral spin textures $\bm{S}^{\text{spiral}}$ would display as a meron-antimeron crystal [Fig.~\ref{fig:fig1}(e)]. The spin chirality of meron and antimeron is opposite and would cancel with each other [Fig.~\ref{fig:fig1}(f)].  In the SkX systems, it is known that the SkX can be tuned into the meron-antimeron crystal via external magnetic fields \cite{Takashi2019, Yu2018}. But the phase shift $\theta$ generally is hard to be continuously tuned in experiments.  In practice, the phase shift $\theta$ in our model can be estimated by mimicking triple-Q CDW and SDW real space patterns in a given experiment.   

\begin{figure}
	\centering
\includegraphics[width=1\linewidth]{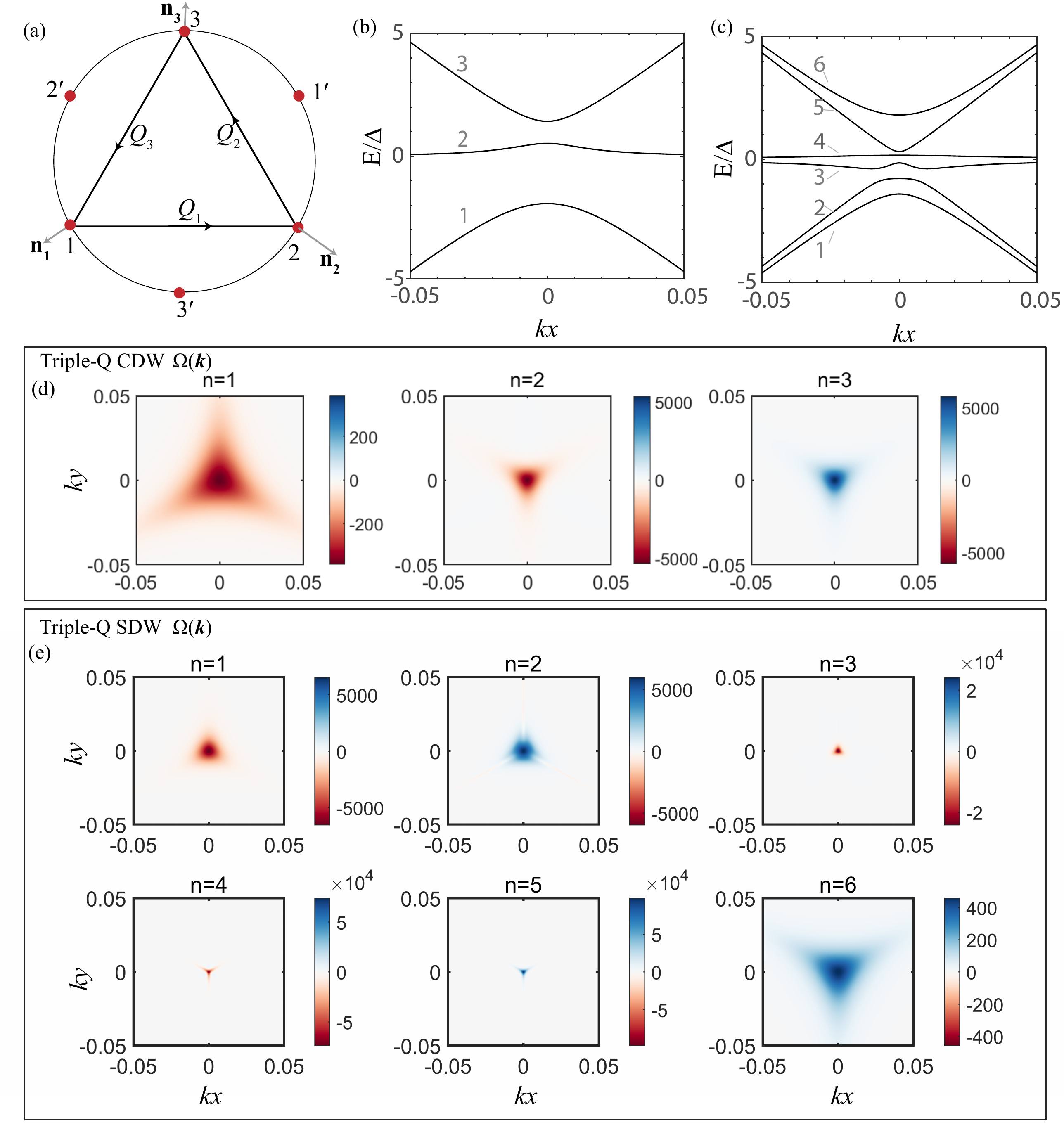}
	\caption{ (a) Illustration of the six `hot spots' (red dots) on the Fermi circle, which are connected through the triple-Q vectors. Note that the dispersion of electrons is given by $\xi_{\bm{p}}=\frac{p^2}{2m}-\mu$, and   $\hat{n}_i$ is simply the unit vector for the direction of the hot-spot momentum $\bm{p}_i$, where the two density wave order parameters interfere. (b) and (c)  respectively, show the band structures of triple-Q CDW and triple-Q SDW along $(k_x,0)$ direction given by the effective Hamiltonians, where the phase shift is set to be $\theta=\pi/4$. (d) and (f) show the Berry curvature $\Omega(\bm{k})$ (in units of $a^2$ and $a$ is the lattice constant) of each band in (b) and (c), respectively.  We set $v_F=100, 
 \Delta=1$ in calculations.    Note that (d) and (e) only show the distribution near one set of hot spots. }
	\label{fig:fig2}
\end{figure}

\emph{Dirac Hamiltonians induced by triple-Q density waves---} Beyond the real-space topology, next we show that the triple-Q density wave order parameters can alter the geometry properties of Bloch bands using a weak-coupling approach. Specifically, we point out that the triple-Q density waves can spontaneously induce Dirac-like physics near the Fermi energy. As the focus of this work is the triple-Q order parameter, we shall set the band dispersion to be the simplest one:  $	\xi_{\bm{p}}=\frac{\bm{p}^2}{2m}-\mu$, where $m$ is an effective mass, $\mu$ is the chemical potential.  In practice, the chemical potential $\mu$ should be located near the Fermi energy where the CDW nesting happens. As shown in Fig.~\ref{fig:fig2}(a), 
there are six `hot spots' (labeled as red dots) that the triple-Q order parameter would couple near Fermi energy. These hot spots can be classified into two sets, i.e., 1,2,3 and 1$'$,2$'$,3$'$. The single-particle dispersion near the hot spots can be approximated as 
\begin{equation}
\epsilon_{\bm{k}}=v_F \bm{\hat{n}}_i\cdot \bm{k},
\end{equation}
where $v_F$ is the Fermi velocity, and unit vectors $\hat{\bm{n}}_1=(-\frac{\sqrt{3}}{2},-\frac{1}{2}), 	\hat{\bm{n}}_2=(\frac{\sqrt{3}}{2},-\frac{1}{2}), 
	\hat{\bm{n}}_3=(0,1)$, and the momentum $\bm{k}$ is measured with respect to the hot-spot momentum $\bm{p}_i$ with $\bm{k}=\bm{p}-\bm{p}_i$.

Next, we present the low-energy effective Hamiltonians with the triple-Q order parameter, which would induce the coupling between these hot spots, i.e., band folding.  Here, we assume the coupling is weak enough so that these hot spots do not overlap with each other. As we will soon see, the additional phase shift $\theta$ modifies the low-energy band dispersions.

We first focus on the triple-Q CDW.
The effective Hamiltonian of the CDW state can be written as $\mathcal{H}=\sum \Phi_{\bm{k}}^{\dagger} H_{\text{CDW}}(\bm{k}) \Phi_{\bm{k}}$ with $\Phi_{\bm{k}}=(c_{1\bm{k}},c_{2\bm{k}}, c_{3\bm{k}})^{T}$ and

\begin{equation}
	H_{\text{CDW}}(\bm{k})=\begin{pmatrix}
		v_F\bm{k}\cdot \hat{\bm{n}}_1& \Delta e^{-i\theta}& \Delta e^{i\theta}\\
		 \Delta e^{i\theta}&v_F\bm{k}\cdot \hat{\bm{n}}_2& \Delta e^{-i\theta}\\
		 \Delta e^{-i\theta}&\Delta e^{i\theta}&v_F\bm{k}\cdot \hat{\bm{n}}_3
	\end{pmatrix},
\end{equation}
where $\bm{k}$ is measured with respect to the hot spots. Note that $\bm{k}=0$ is the Gamma point of the folded Brillouin zone.  The effective Hamiltonian $H'_{\text{CDW}}(\bm{k})$ defined  in the other group $(c_{1'\bm{k}},c_{2'\bm{k}}, c_{3'\bm{k}})^{T}$ can be obtained by replacing $\hat{\bm{n}}_i$ and $\theta$ in $H_{\text{CDW}}(\bm{k})$ with $\hat{\bm{n}'}_i=-\hat{\bm{n}}_i$ and $-\theta$.

To emphasize the nontrivial band geometry encoded in the effective Hamiltonian, we show that $H_{\text{CDW}}(\bm{k})$ can be actually mapped to a three-band Dirac Hamiltonian. Note that the multi-band Dirac Hamiltonian in this work is defined as all non-vanishing terms in the Hamiltonian are linearly proportional to $k_{+}$ or $k_{-}$\cite{Law2022}. As shown in Supplementary Material (SM) \cite{Supp}, the eigenstates of $H_{\text{CDW}}(\bm{k})$ at $\bm{k}=0$ can be labeled with the eigenvalue of $C_{3z}$ operation: $C_{3z}\ket{E_{J,m}}=e^{-i\frac{2m\pi}{3}}\ket{E_{J,m}}$,  where the total angular momentum $J=1$, the magnetic angular momentum $m=0,\pm1$. After projecting the $H_{\text{CDW}}(\bm{k})$ into the `good' basis spanned by $(\ket{E_{1,0}},\ket{E_{1,-1}},\ket{E_{1,1}})$, the effective Hamiltonian is represented as a three-band Dirac Hamiltonian:
\begin{equation}
H_{\text{CDW}}^{\text{eff}}(\bm{k})=\begin{pmatrix}
E_{1,0} &-\frac{i}{2}v_Fk_{-}&-\frac{i}{2}v_Fk_{+}\\
\frac{i}{2}v_Fk_{+}&		E_{1,-1}&-\frac{i}{2}v_Fk_{-}\\
\frac{i}{2}v_Fk_{-}&	\frac{i}{2}v_Fk_{+}&E_{1,1}
	\end{pmatrix}
\end{equation}
where $k_{\pm}=k_x\pm ik_y$, $E_{J,m}=2\Delta\cos(\theta-\frac{2m\pi}{3})$. Importantly, the phase shift plays a significant role in affecting the Dirac mass.   It is also noted that the multi-band Dirac Hamiltonian can also be induced in double-Q when more than two hot spots are folded together simultaneously (see SM \cite{Supp} for details). We shall focus on the triple-Q density waves case below as the phase shift degrees of freedom in single-Q and double-Q ones are absent.

Similar physics can also be induced by the triple-Q SDW order parameters. Specifically, we can first obtain the effective Hamiltonian near the first set of hot spots (1,2, and 3) as
  \begin{equation}
  	H_{\text{SDW}}(\bm{k})=\begin{pmatrix}
  		v_F \bm{k}\cdot \hat{\bm{n}}_1 & \Delta  \bm{S}_{-\bm{Q}_1}\cdot \bm{\sigma}& \Delta  \bm{S}_{\bm{Q}_3}\cdot \bm{\sigma}\\
  \Delta	\bm{S}_{\bm{Q}_1}\cdot \bm{\sigma}& 		v_F \bm{k}\cdot \hat{\bm{n}}_2&   	\Delta \bm{S}_{-\bm{Q}_2}\cdot \bm{\sigma}\\
  	 \Delta  \bm{S}_{-\bm{Q}_3}\cdot \bm{\sigma}&   	 \Delta  \bm{S}_{\bm{Q}_2}\cdot \bm{\sigma}& v_F\bm{k}\cdot \hat{\bm{n}}_3
  	\end{pmatrix},
  \end{equation}
where the basis is $(c_{1\bm{k}}, c_{2\bm{k}}, c_{3\bm{k}} )^{T}\otimes
(\uparrow,\downarrow)^{T}$, and 	$\bm{S}_{\pm \bm{Q}_\nu}=(\mp\frac{i}{2} \cos \phi_{\nu}, \mp\frac{i}{2}\sin \phi_{\nu}, \mp \frac{1}{2}) e^{i\pm \theta}$. By replacing $\bm{S}_{\pm \bm{Q}_\nu}$ with $\bm{S}_{\mp \bm{Q}_\nu}$ and $\hat{\bm{n}}_i$ with $\hat{\bm{n}}'_i$,  the low-energy Hamiltonian $H'_{\text{SDW}}(\bm{k})$ near the other set of hot spots  ($1'$, $2'$, $3'$) can be obtained.  Also,  performing a unitary transform and rewriting the Hamiltonian in the basis $\ket{E}_{J,m}\otimes \{\uparrow, \downarrow\}$, a six-band effective Hamiltonian can be obtained
(see
SM Note 1 for the detailed form).

One important property of the Dirac Hamiltonian is to enable the Berry curvature, which has been associated with various topological materials. Here, we show that the triple-Q density wave order parameters can induce Berry curvature on the simplest quadratic band.   The band structures given by $H_{\text{CDW}}(\bm{k})$ and $H_{\text{SDW}}(\bm{k})$ are shown Figs.~\ref{fig:fig2}(b) and \ref{fig:fig2}(c), respectively.  The Berry curvature of these bands can be easily calculated, as shown in Fig.~\ref{fig:fig2}(d) for the triple-Q CDW and Fig.~\ref{fig:fig2}(e)  for the triple-Q SDW.  Moreover, we find that the phase shift would affect the Berry curvature significantly. For example, the Berry curvature vanishes at $\theta=\frac{n\pi}{3}$ and $\theta=\frac{(2n+1)\pi}{6}$ for the $H_{\text{CDW}}(\bm{k})$ and $H_{\text{SDW}}(\bm{k})$ respectively, while the latter exactly corresponds to the presence of meron-antimron crystals in real space.  %Note that the Berry curvature given by $H'(\bm{k})$ and $H'_{SDW}(\bm{k})$. 

 \begin{table}
 	\caption{Possible responses induced triple-Q density wave order parameters from symmetry principle. Here,  AHE, SHG, MOE, VHE label the anomalous Hall effect, second harmonic generation, magnetoptical effect, valley Hall effect.  }
 	\begin{tabular}{|c|c|c|c|}
 	\hline\hline
 		Triple-Q & 	$\theta$ & key symmetries &  responses  \\\hline
	\multirow{2}{*}{CDW}	 
 &$\theta=\frac{n\pi}{3}$& $T$,  $I$, $C_{3z}$, $M_x$, $M_y$ &\\    \cline{2-4}
 		& $\theta\neq \frac{n\pi}{3}$& $T$,$\cancel{I}$, $C_{3z}$, $\cancel{M_x}$, $M_y$&  SHG, VHE\\\hline
 \multirow{2}{*}{SDW}& $\theta=\frac{n\pi}{3}$&$\cancel{T}$,$\cancel{I}$, $C_{2z}$, $C_{3z}$  & AHE, MOE\\\cline{2-4}
 &$\theta\neq \frac{n\pi}{3}$ & $\cancel{T}$,$\cancel{I}$, $\cancel{C_{2z}}$, $C_{3z}$ & AHE, MOE SHG \\\hline\hline
 	\end{tabular}
  \label{tab3}
 \end{table}

\begin{figure}
	\centering
\includegraphics[width=1\linewidth]{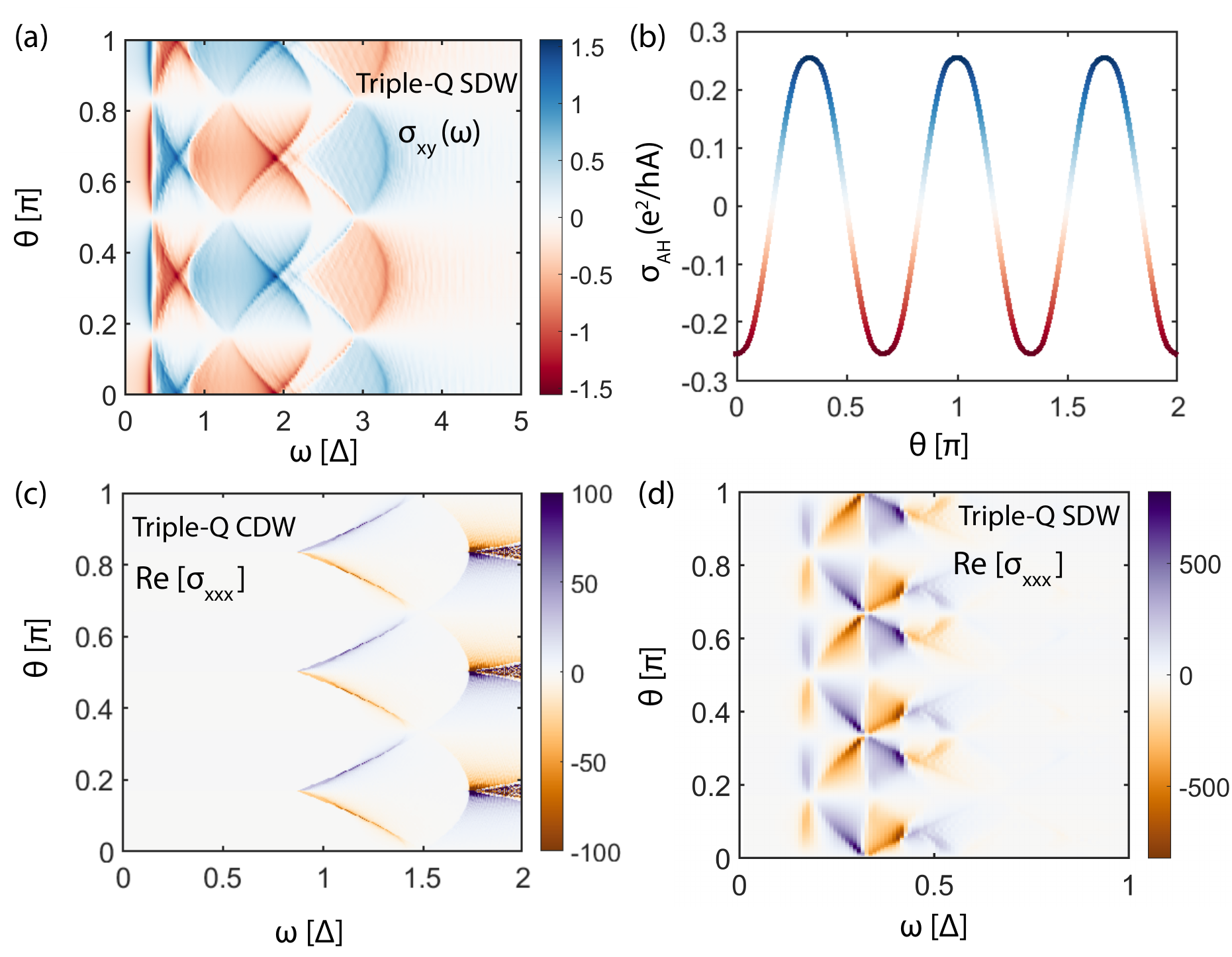}
	\caption{(a) The magnetoptical conductivity $\sigma_{xy}(\omega)$ as a function of the phase shift $\theta$ and the photon frequency $\omega$, where the colorbar is in unit of $\frac{e^2}{h}$. (b) The anomalous Hall conductivity as a function of photon frequency.  (c) and (d), respectively, show the real part of the second-harmonic response tensor $\text{Re}[\sigma_{xxx}(2\omega;\omega,\omega)]$, which is in units of $e^3/(2\hbar^2)$. The chemical potential in this figure is fixed at the position of hot spots, and the temperature is zero temperature limit with $T=0.01\Delta$. }
	\label{fig:fig3}
\end{figure}

\emph{Linear and nonlinear responses in triple-Q density wave systems---}
Due to the presence of nontrivial band geometry, we expect that triple-Q density wave order parameters would naturally result in some interesting linear and nonlinear responses. Let us identify the symmetries dedicated by the triple-Q density wave order parameter first.  Either from the real-space landscape or the effective Hamiltonian  (see SM \cite{Supp} for detailed symmetry analysis), we can identify that the triple-Q CDW order parameter $\Delta_{\text{CDW}}(\bm{r})$  preserves time-reversal symmetry at generic $\theta$  but breaks inversion only when $\theta\neq \frac{n\pi}{3}$. Consequently, a second harmonic generation (SHG) solely induced by the triple-Q CDW would be expected.  Furthermore, if we consider the two sets of hotspots within CDWs as distinct valleys within momentum space, the Berry curvature contrasts between these valleys due to time-reversal symmetry. This contrast can induce valley Hall effects \cite{Sui2015,Shimazaki2015,Qianniu2008,Jianbo}.   

On the other hand, it can be shown that the triple-Q SDW order parameter $\Delta_{\text{SDW}}(\bm{r})$ would break both time-reversal and inversion symmetry regardless of the value of $\theta$, which stems from the presence of spiral spin textures.    The broken time-reversal symmetry allows the anomalous Hall effect (AHE) and magnetoptical responses, while the further broken inversion symmetry is crucial for the nonreciprocal or nonlinear responses \cite{Tokura2018}. The resulting key point group symmetry generators induced by the $\Delta_{\text{SDW}}(\bm{r})$ are  $C_{2z}$ at $\theta=\frac{n\pi}{3}$, and  $C_{2z}$ would be broken when $\theta\neq \frac{n\pi}{3}$. Note that, unlike triple-Q CDWs, the point group formed by the unitary operations of triple-Q SDWs is chiral, where all mirror symmetries are broken.  As a result,  the nonreciprocal nonlinear responses can be supported in triple-Q SDW systems as well.   To be clear, the above discussions are explicitly summarized in Table~\ref{tab3}.    Note that as imposed by us, the $C_{3z}$ symmetry is also preserved at generic $\theta$ for triple-Q density waves in this work. 
 When the $C_{3z}$ symmetry is absent, such as broken by the external strain, other interesting effects may further be allowed (see SM Note 3 for a discussion of nonlinear Hall effects \cite{Liang2015,Du2021,Ma2019} in strained SDW states).

%\begin{equation}
    	%\sigma_{xy}(\omega)=\hbar e^2 \int \frac{d^2\bm{k}}{(2\pi)^2}\sum_{n\neq n'}\frac{(f_{n\bm{k}}-f_{n'\bm{k}}) \text{Im} [v_{nn'}^{x}(\bm{k})v_{nn'}^{y}(\bm{k})]}{(\epsilon_{n\bm{k}}-\epsilon_{n'\bm{k}})^2-(\hbar\omega+i\eta)^2}.
    %\end{equation}
    %Here, $\omega$ is the photon frequency, $f_{n\bm{k}}$ is the Fermi distribution function for the $n$-th band $\epsilon_{n\bm{k}}$, the veolcity function $v^{\alpha}_{nn'}=\braket{n\bm{k}|\frac{\partial H}{\partial \bm{k}_{\alpha}}|n\bm{k}}$. Note that the anomalous Hall effect is given by the DC response $\sigma_{\text{AHE}}=\sigma(\omega=0)$.
    To be more explicit, we now calculate some representative linear and nonlinear responses from the effective triple-Q CDW and SDW Hamiltonians. The representative linear responses: AHE and magnetoptical conductivity can be obtained from the Kubo formula \cite{Feng2020}. Using the triple-Q effective Hamiltonian $H_{\text{SDW}}(\bm{k})$ and $H'_{\text{SDW}}(\bm{k})$ (these two contributions need to be summed up),  the calculated magnetoptical conductivity $\sigma_{xy}(\omega)$  as a function the frequency $\omega$ and the phase shift $\theta$, and the calculated AHE conductivity  $\sigma_{\text{AH}}$ versus $\theta$ are shown in Fig.~\ref{fig:fig3} (a) and (b), respectively. The finite AHE and magnetoptical effect are expected due to the nontrivial band geometry near Fermi energy. It is worth noting that the AHE vanishes at $\theta=\frac{(2n+1)\pi}{6}$, while remaining finite at other phase shifts. Such observation is consistent with the real-space topology discussed in Fig.~\ref{fig:fig1}: the meron-antimeron crystal appears at $\theta=\frac{(2n+1)\pi}{6}$, while the skyrmion crystals that possess topological Hall effects would appear at other phase shifts. 

Finally, we highlight the representative nonlinear responses enabled by the triple-Q density wave order parameters. Nonlinear responses have been a subject of intense study in recent years (see reviews \cite{Tokura2018, Morimoto2023, Yanase2024} for experimental and theoretical details). Observing these nonlinear responses requires certain symmetry conditions. For instance, SHG necessitates inversion symmetry breaking, while magnetochiral anisotropy requires the breaking of both inversion and time-reversal symmetry. Additionally, band geometry plays a significant role in enhancing nonlinear responses as well \cite{Morimoto2016, Morimoto2023}. Guided by these principles, we anticipate that a triple-Q density wave state could be highly effective in exploring nonlinear responses due to its involvement in both symmetry breaking and nontrivial band geometry.   As a demonstration, the second harmonic generation response tensor $\sigma_{xxx}(2\omega; 
\omega, \omega)$ calculated from the effective Hamiltonians of triple-Q CDW and SDW are shown in Fig.~\ref{fig:fig3}(c) to (d) (see the explicit SHG formalism in  refs.~\cite{Moore2019, Takasan2021} or in Supplementary Note 2). Remarkably, the SHG is finite at a wide parameter region.   Some special phase shifts where the SHG vanishes also match with the expectations: (i) $\theta=\frac{2n\pi}{3}$ for triple-Q CDW due to the restoration of inversion symmetry, (ii) $\theta=\frac{n\pi}{3}$ and $\theta=\frac{(2n+1)\pi}{6}$ for triple-Q CDW and SDW, where the inversion symmetry is still broken but the Berry curvature vanishes. Additionally, as presented in Supplementary Information Note 1,  we also can show that the magnetochiral anisotropy would be supported in tripe-Q  SDW at $\theta\neq \frac{n\pi}{3}$ with $C_{2z}$ symmetry broken.

\emph{Discussion---} It is noteworthy that various intriguing linear and nonlinear responses have been experimentally studied in several triple-Q density wave systems. For example, it was observed that the inversion and mirror symmetry are broken in the triple-Q CDW system  1T TiSe$_2$ \cite{Fichera2020}, which are indeed consistent with our analysis in Table II of triple-Q CDW with finite phase shifts $\theta\neq \frac{n\pi}{3}$.  On the other hand, it was suggested that there are both SkX (called A phase) and meron-antimeron crystal (called IC-1 phase) in Gd$_2$PdSi$_3$\cite{Takashi2019}. In ref.~\cite{Kato2023}, the magnetoptical Kerr effect (MOKE)  was experimentally studied in Gd$_2$PdSi$_3$. Indeed, the SkX region displays much more salient MOKE than the meron-antimeron crystal, which is also aligned with our analysis in Fig.~\ref{fig:fig3}(a). Additionally, the chiral transport effect was observed in Kagome triple-Q CDW  metal CsV$_3$Sb$_5$ and indicates time-reversal symmetry breaking \cite{Guo2022}. To capture such features, we believe that the multi-sublattice degree of freedom in CsV$_3$Sb$_5$ plays an important role and results in complex triple-Q order parameters. How the phase shift plays a role in affecting the complex triple-Q CDW order parameters remains an interesting future problem.

  It is interesting to note that for the triple-Q CDW, the inversion can only be broken with a non-vanishing phase shift (see Table II). Motivated by this observation, we propose to measure the SHG in the triple-Q CDW, where the parent crystal structure is centrosymmetric but the triple-Q CDW spontaneously breaks the inversion.  In this case, the SHG signal arises from the inversion symmetry breaking of triple-Q CDW, which is directly associated with the presence of phase shifts. Although a related experiment has been conducted in 1T TiSe2 \cite{Fichera2020}, the role of phase shift has not yet been emphasized. Moreover, the observation of other responses listed in Table II, such as the valley Hall effects, would further verify our theory.

\begin{comment}
  \textcolor{blue}{\textcolor{red}{In conclusion, we have provided a minimal but comprehensive theory to study the interference of multi-Q density waves on electronic states. } 
 Our theoretical work significantly advances the understanding of phase shifts, band geometry, and the resulting linear and nonlinear responses in multi-Q density waves.}  
\end{comment}

%It can also be seen  that the SHG signal of triple-Q SDW typically is several times larger than the triple-Q CDW. 
%\YM{Emphasize that only  the terms associated with the Berry connection contribute} \YM{Inserting some real number and estimating how large}\YM{Metallic states negalectable}

%\emph{Conclusions.---} In conclusion, we have provided a minimal but comprehensive theory to capture the important physical consequences induced by both triple-Q CDW and SDW order parameters.   
%Our theoretical work significantly advances the understanding of phase shifts, band geometry, and the resulting linear and nonlinear responses in triple-Q density waves. %The nontrivial band geometry in density waves . 

\emph{ Acknowledgements---}We are grateful to Hiroki Isobe and  Wataru Koshibae for enlightening discussions.
N.N. was supported by JSPS KAKENHI Grant No. 24H00197 and 24H02231.
N.N. was also supported by the RIKEN TRIP initiative.  Y.M.X.  acknowledges financial support from the RIKEN Special Postdoctoral Researcher(SPDR) Program.

 \vspace{1\baselineskip}

\clearpage
	
	\onecolumngrid
	\begin{center}
		\textbf{\large Supplementary Material for  ``Phase shifts, band geometry and responses in triple-Q charge and spin density waves'' }\\[.2cm]
		Ying-Ming Xie $^{1}$  and Naoto Nagaosa$^{1,2}$ \\[.1cm]
		{\itshape ${}^1$    RIKEN Center for Emergent Matter Science (CEMS), Wako, Saitama 351-0198, Japan}\\[.1cm]
			{\itshape ${}^2$Fundamental Quantum Science Program, TRIP Headquarters, RIKEN, Wako 351-0198, Japan}\\[1cm]
	\end{center}
	\setcounter{equation}{0}
\setcounter{section}{0}
\setcounter{figure}{0}
\setcounter{table}{0}
\setcounter{page}{1}
\renewcommand{\theequation}{S\arabic{equation}}
\renewcommand{\thesection}{Supplementary Note \arabic{section}}
 \renewcommand{\figurename}{Supplementary Figure}
 \renewcommand{\thetable}{\arabic{table}}
 \renewcommand{\tablename}{Supplementary Table}
	\makeatletter
	
	\onecolumngrid
	
	\maketitle

\section{Effective Hamiltonian 
 and symmetry analysis for Triple-Q CDW state}

\subsection{ Effective Hamiltonian and symmetry analysis for the Triple-Q CDW state.} 

As shown in Supplementary Fig.~\ref{fig:figs1} (a) and (b), there are six hot spots connected by the triple-Q vectors. These hot spots can be classified into two groups, i.e., 1,2,3 and 1$'$,2$'$,3$'$. The dispersion $	\xi_{\bm{k}}=\frac{\hbar^2\bm{k}^2}{2m}-\mu$ near the hot spots can be approximated as 
\begin{equation}
\xi_{\bm{k}}=v_F \bm{\hat{n}}_i\cdot \bm{k},
\end{equation}
where $v_F$ is the Fermi velocity, and 
\begin{equation}
	\hat{\bm{n}}_1=(-\frac{\sqrt{3}}{2},-\frac{1}{2}), 	\hat{\bm{n}}_2=(\frac{\sqrt{3}}{2},-\frac{1}{2}), 
	\hat{\bm{n}}_3=(0,1).
\end{equation}
In real space, the CDW order parameters can be written as
\begin{equation}
	\Delta (\bm{r}_i)=2\Delta\sum_{\nu}\cos(\bm{Q}_{\nu}\cdot\bm{r}_i+\theta_{\nu}).
\end{equation}
Here,
\begin{equation}
	\bm{Q}_1=(Q,0), \bm{Q}_2=(-Q/2,\sqrt{3}Q/2), \bm{Q}_3=(-Q/2, -\sqrt{3}Q/2).
\end{equation} 

The effective Hamiltonian of the CDW state can be written as $\mathcal{H}=\sum \Phi_{\bm{k}}^{\dagger} H_{\text{CDW}}(\bm{k}) \Phi_{\bm{k}}$ with $\Phi_{\bm{k}}=(c_{1\bm{k}},c_{2\bm{k}}, c_{3\bm{k}})^{T}$ and

\begin{equation}
	H_{\text{CDW}}(\bm{k})=\begin{pmatrix}
		v_F\bm{k}\cdot \hat{\bm{n}}_1& \Delta e^{-i\theta_1}& \Delta e^{i\theta_3}\\
		 \Delta e^{i\theta_1}&v_F\bm{k}\cdot \hat{\bm{n}}_2& \Delta e^{-i\theta_2}\\
		 \Delta e^{-i\theta_3}&\Delta e^{i\theta_2}&v_F\bm{k}\cdot \hat{\bm{n}}_3
	\end{pmatrix},
\end{equation}
where $\bm{k}$ is measured respected from $\bm{p}_i$ the unit vectors in triple-Q case are $\hat{\bm{n}}_1=(-\frac{\sqrt{3}}{2},-\frac{1}{2}), 	\hat{\bm{n}}_2=(\frac{\sqrt{3}}{2},-\frac{1}{2}), 
	\hat{\bm{n}}_3=(0,1)$.  Near the other group hot spots, the effective Hamiltonian becomes
\begin{equation}
	H'_{\text{CDW}}(\bm{k})=\begin{pmatrix}
		v_F\bm{k}\cdot \hat{\bm{n}}'_1& \Delta e^{i\theta_1}& \Delta e^{-i\theta_3}\\
		\Delta e^{-i\theta_1}&v_F\bm{k}\cdot \hat{\bm{n}}'_2& \Delta e^{i\theta_2}\\
		\Delta e^{i\theta_3}&\Delta e^{-i\theta_2}&v_F\bm{k}\cdot \hat{\bm{n}}'_3
	\end{pmatrix}.
\end{equation}

Next, let us deduce the symmetry operations for the effective Hamiltonian.  First of all, it is easy to show that the system respects the spinless time-reversal symmetry $\mathcal{T}=K$ ($K$ the conjugate)
\begin{equation}
	\mathcal{T} 	H'_{\text{CDW}}(-\bm{k})\mathcal{T}^{-1}=H_{\text{CDW}}(\bm{k}).
\end{equation}
The representation of out-of-plane three-fold rotation  can be written as
\begin{equation}
	U_{C_3}=\begin{pmatrix}
		0&1&0\\
		0&0&1\\
		1&0&0
	\end{pmatrix}.
\end{equation}
Then it is straightforward to show that under the $C_3$ operation,
\begin{equation}
	U_{C_3}H_{\text{CDW}}(C_3\bm{k})	U_{C_3}^{-1}=\begin{pmatrix}
			v_F\bm{k}\cdot \hat{\bm{n}}_1& \Delta e^{-i\theta_2}& \Delta e^{i\theta_1}\\
			\Delta e^{i\theta_2}&v_F\bm{k}\cdot \hat{\bm{n}}_2& \Delta e^{-i\theta_3}\\
			\Delta e^{-i\theta_1}&\Delta e^{i\theta_3}&v_F\bm{k}\cdot \hat{\bm{n}}_3
		\end{pmatrix}.
\end{equation}
We can see that the system  preserves $C_3$ symmetry $	U_{C_3}H_{\text{CDW}}(C_3\bm{k})	U_{C_3}^{-1}=H_{\text{CDW}}(C_3\bm{k})$ if $\theta_1=\theta_2=\theta_3=\theta$.
Similarly, $	U_{C_3}H'_{\text{CDW}}(C_3\bm{k})	U_{C_3}^{-1}=H'_{\text{CDW}}(C_3\bm{k})$ so that the full system respects $C_3$ symmetry. Hence, we would fix the $\theta_1=\theta_2=\theta_3$ in our discussion.

It can be found that when $\theta=\frac{n\pi}{3}$ ($n$ are integers), the system further respects the inversion symmetry and $M_x$:
\begin{eqnarray}
	I H'_{\text{CDW}}(-\bm{k})I^{-1}&&=H_{\text{CDW}}(\bm{k}), \label{Eq_10}\\
 M_xH_{\text{CDW}}(-k_x,k_y)M_x^{-1}&&= H_{\text{CDW}}(k_x,k_y).
\end{eqnarray}

Let us show the above two equations in more detail. The mirror $M_x$ operation maps $k_x$  to $-k_x$, and exchanges hot spots 1 and 2 within the same group. As a result,
\begin{eqnarray}
M_xH_{\text{CDW}}(-k_x,k_y)M_x^{-1}&&=\begin{pmatrix}
0&1&0\\
1&0&0\\
0&0&1
\end{pmatrix}\begin{pmatrix}
v_F(\frac{\sqrt{3}}{2}k_x-\frac{1}{2}k_y) &\Delta e^{-i\theta}&\Delta e^{i\theta}\\
\Delta e^{i\theta}& v_F(-\frac{\sqrt{3}}{2}k_x-\frac{1}{2}k_y)&\Delta e^{-i\theta}\\
\Delta e^{-i\theta}&\Delta e^{i\theta}& v_F k_y
\end{pmatrix}\begin{pmatrix}
    0&1&0\\
    1&0&0\\
    0&0&1
\end{pmatrix}\nonumber\\
&&=\begin{pmatrix}
    v_F\bm{k}\cdot\hat{\bm{n}}_1 &\Delta e^{i\theta} &\Delta e^{-i\theta}\\
    \Delta e^{-i\theta}& v_F\bm{k}\cdot\hat{\bm{n}}_2&\Delta e^{i\theta}\\
    \Delta e^{i\theta}&\Delta e^{-i\theta}& v_F\bm{k}\cdot \bm{\hat{n}}_3 \label{Eq_S12}
\end{pmatrix}.
\end{eqnarray}
When $\theta=-\theta$, i.e., $\theta=n\pi$, Eq.~\eqref{Eq_S12} becomes $H_{\text{CDW}}(\bm{k})$. It is also worth noting that when $\theta_1+\theta_2+\theta_3=3\theta=2n\pi$, the ground states are the same \cite{Fukuyama2014}. In other words, the system exhibits a periodicity of $\frac{2\pi}{3}$. As a result, when $\theta=n\pi/3$, we generally expect 
 $M_xH_{\text{CDW}}(-k_x,k_y)M_x^{-1}= H_{\text{CDW}}(k_x,k_y)$, i.e., the presence of mirror symmetry. Similarly, the relation Eq.~\eqref{Eq_10} can be shown.
Note that the effects of inversion and out-of-plane $C_2$ are the same here, which maps momentum $\bm{k}$ to $-\bm{k}$, and maps one group hot spots to the other group.

Finally, 
for all $\theta$, the system exhibits $M_y$ symmetry:
\begin{equation}
M_yH_{\text{CDW}}(k_x,-k_y)	M_y^{-1}=H'_{\text{CDW}}(k_x,k_y).
\end{equation}

 \begin{figure}
 	\centering
 	\includegraphics[width=0.7\linewidth]{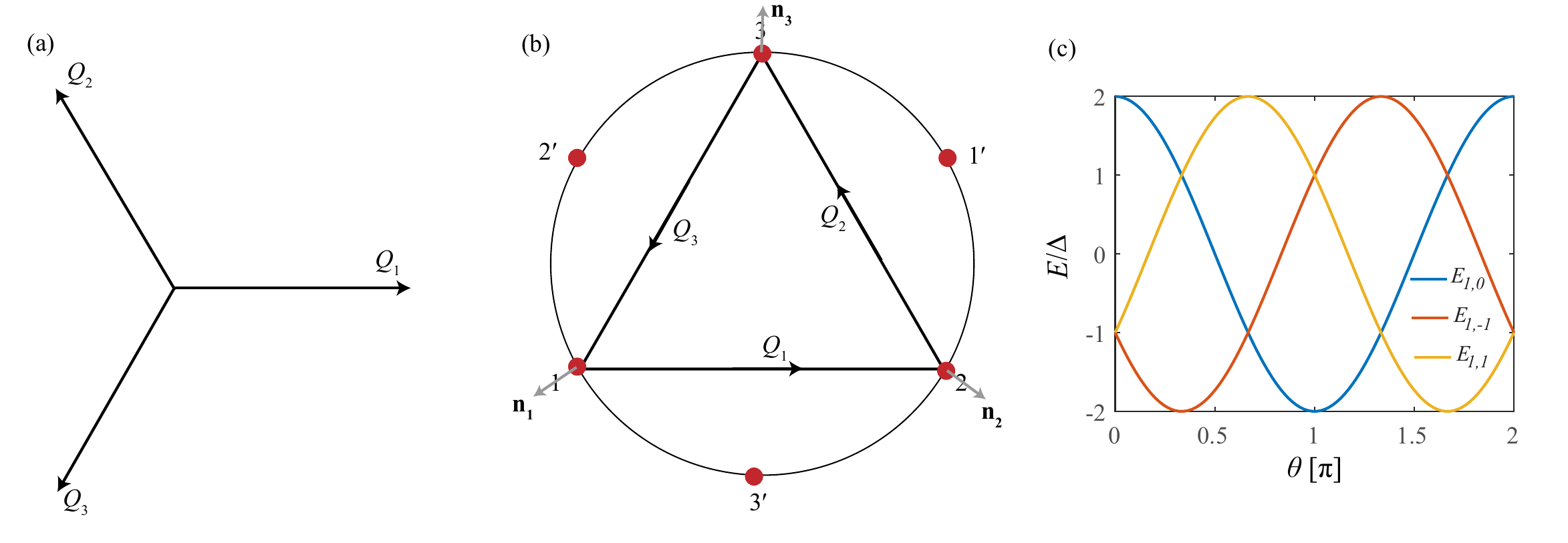}
 	\caption{(a) Triple Q vectors: $\bm{Q}_1$, $\bm{Q}_2$, $\bm{Q}_3$. (b) Six hot spots in k-space, where the gap opening due to density waves overlap. These six hot spots can be classified into two groups:(1,2,3) and (1$'$,2$'$,3$'$). (c) The energy $E_{J,m}$ versus $\theta$ with $J=1$, $m=0,\pm 1$. }
 	\label{fig:figs1}
 \end{figure}

In the main text, we have mapped $H_{\text{CDW}}(\bm{k})$ as a three-band massive Dirac Hamiltonian.  Now we present the details here. 
We can solve the eigenstates of $H(\bm{k})$ analytically  
at the hot spots ($\bm{k}=0$): 
\begin{equation}
\begin{split}
E_{1,0}&=2\Delta \cos\theta,\\  
E_{1,-1}&=2\Delta\cos(\frac{2\pi}{3}+\theta), \\
E_{1,1}&=2\Delta\cos(\frac{4\pi}{3}+\theta).
\end{split}
\end{equation}
See Supplementary Fig.~1 for a plot of these three energy levels versus $\theta$. The corresponding eigenstates are
\begin{equation}
\begin{split}
\ket{E_{1,0}}&=\frac{1}{\sqrt{3}}(\ket{p_1}+\ket{p_2}+\ket{p_3}),
 \\
\ket{E_{1,-1}}&=\frac{1}{\sqrt{3}}(e^{i\frac{\pi}{3}}\ket{p_1}+e^{-i\frac{\pi}{3}}\ket{p_2}+e^{i\pi}\ket{p_3}),\\
\ket{E_{1,1}}&=\frac{1}{\sqrt{3}}(e^{i\frac{2\pi}{3}}\ket{p_1}+e^{i\frac{4\pi}{3}}\ket{p_2}+\ket{p_3}).
\end{split}
\end{equation}
where $\ket{p_j}$ labels the wavefunction of $j$-th hot spots, the eigenstates as $\ket{E_{J,m}}$ with $J$ as the total angular momentum number, $m$ as the magnetic quantum number.  It is easy to verify that  $C_{3}\ket{E_{J,m}}=e^{-i\frac{2m\pi}{3}}\ket{E_{J,m}}$  under the splinless $C_{3}$ operation. 
Then we can project the Hamiltonian in the basis spanned by $(\ket{E_{1,0}},\ket{E_{1,-1}},\ket{E_{1,1}})$:
\begin{equation}
H^{\text{eff}}_{\text{CDW}}(\bm{k})=\begin{pmatrix}
E_{1,0} &-\frac{i}{2}v_Fk_{-}&-\frac{i}{2}v_Fk_{+}\\
\frac{i}{2}v_Fk_{+}&		E_{1,-1}&-\frac{i}{2}v_Fk_{-}\\
\frac{i}{2}v_Fk_{-}&	\frac{i}{2}v_Fk_{+}&E_{1,1}
	\end{pmatrix}
\end{equation}

%It can be seen that Eq.~\eqref{Eq_S15} is equivalent to the main text Eq.~(7). %It is worth noting that this three-band Dirac Hamiltonian was also observed in the moir\'e materials with triple-Q moir\'e potential .

\subsection{Effective Hamiltonian and symmetry analysis for the Triple-Q SDW state}

As mentioned in the main text, we can express the SDW states with the superposition of spiral spin textures
\begin{equation}
	\bm{S}_i^{spiral}=\Delta \sum_{\nu=1}^{3} (\sin \mathcal{Q}_{\nu} \cos \phi_{\nu}, \sin \mathcal{Q}_{\nu} \sin \phi_{\nu}, -\cos \mathcal{Q}_{\nu}).
\end{equation}
where $i$ is the site index, $\phi_{\nu}=\frac{2}{3}\pi (\nu-1)$, $\mathcal{Q}_{\nu}=\bm{Q}_{\nu}\cdot \bm{r}_i+\theta_{\nu}$.

  It is easy to verify that the spin Hamiltonian caused by the SDW states is $C_3$ invariant:
 \begin{equation}
 	H_s=\sum_{i} \bm{S}_i^{spiral}\cdot \bm{\sigma}.
 \end{equation}
Then, we can obtain the effective Hamiltonian near the first group hot spots (1,2, and 3) as
  \begin{equation}
  	H_{\text{SDW}}(\bm{k})=\begin{pmatrix}
  		v_F \bm{k}\cdot \bm{n}_1 & \Delta  \bm{S}_{-\bm{Q}_1}\cdot \bm{\sigma}& \Delta  \bm{S}_{\bm{Q}_3}\cdot \bm{\sigma}\\
  \Delta	\bm{S}_{\bm{Q}_1}\cdot \bm{\sigma}& 		v_F \bm{k}\cdot \bm{n}_2&   	\Delta \bm{S}_{-\bm{Q}_2}\cdot \bm{\sigma}\\
  	 \Delta  \bm{S}_{-\bm{Q}_3}\cdot \bm{\sigma}&   	 \Delta  \bm{S}_{\bm{Q}_2}\cdot \bm{\sigma}& v_F\bm{k}\cdot \bm{n}_3
  	\end{pmatrix}.
  \end{equation}
   where the basis is $(c_{1\bm{k},\uparrow},c_{1\bm{k},\downarrow},c_{2\bm{k},\uparrow}, c_{2\bm{k},\downarrow}, c_{3\bm{k},\uparrow} , c_{3\bm{k},\downarrow} )^{T}$, and
   \begin{eqnarray}
   	\bm{S}_{\bm{Q}_\nu}&&=(-\frac{i}{2} \cos \phi_{\nu}, -\frac{i}{2}\sin \phi_{\nu}, -\frac{1}{2}) e^{i\theta_{\nu}},\\
   	   	\bm{S}_{-\bm{Q}_\nu}&&=(\frac{i}{2} \cos \phi_{\nu}, \frac{i}{2}\sin \phi_{\nu}, -\frac{1}{2})e^{-i\theta_{\nu}}.
   \end{eqnarray}
   Near the second group hot spots (1$'$, 2$'$, and 3$'$), the effective Hamiltonian is
    \begin{equation}
     	H'_{\text{SDW}}(\bm{k})=\begin{pmatrix}
    		v_F \bm{k}\cdot \bm{n}'_1 & \Delta  \bm{S}_{\bm{Q}_1}\cdot \bm{\sigma}& \Delta  \bm{S}_{-\bm{Q}_3}\cdot \bm{\sigma}\\
    		\Delta	\bm{S}_{-\bm{Q}_1}\cdot \bm{\sigma}& 		v_F \bm{k}\cdot \bm{n}'_2&   	\Delta \bm{S}_{\bm{Q}_2}\cdot \bm{\sigma}\\
    		\Delta  \bm{S}_{\bm{Q}_3}\cdot \bm{\sigma}&   	 \Delta  \bm{S}_{-\bm{Q}_2}\cdot \bm{\sigma}& v_F\bm{k}\cdot \bm{n}'_3
    	\end{pmatrix}
    \end{equation} 
Again, we can easily show that the system respects the out-of-plane $C_{3z}$ symmetry if  $\theta=\theta_1=\theta_2=\theta_3$ with 
    \begin{eqnarray}
    	U_{C_{3z}}H_{\text{SDW}}(C_3\bm{k})	U_{C_{3z}}^{-1}&&=H_{\text{SDW}}(\bm{k}),  \nonumber\\	U_{C_{3z}}H'_{\text{SDW}}(C_3\bm{k})	U_{C_{3z}}^{-1}&&=H'_{\text{SDW}}(\bm{k}),
    \end{eqnarray}
     where the representation of $C_3$ operation is 
    \begin{equation}
    	U_{C_3}=\begin{pmatrix}
    		0&1&0\\
    		0&0&1\\
    		1&0&0
    	\end{pmatrix}\otimes e^{-i\frac{\pi}{3}\sigma_z}.
    \end{equation}  
     It can also be seen that the system breaks the time-reversal symmetry $\hat{T}=i\sigma_yK$, with
    \begin{equation}
   	  \hat{T} H'_{\text{SDW}}(-\bm{k}) \hat{T}^{-1}\neq 	H_{\text{SDW}}(\bm{k}) .
    \end{equation}
    The inversion symmetry is also broken in general:
 \begin{equation}
 	\hat{I} H'_{\text{SDW}}(-\bm{k})\hat{I}^{-1}\neq H_{\text{SDW}}(\bm{k}).
 \end{equation}
    But there is an emergent `particle-hole'  symmetry:
   \begin{equation}
   	H_{\text{SDW}}(\bm{k})=-\hat{T} H'_{SDW}(\bm{k})\hat{T}^{-1},
   \end{equation} 
    and it further respects the combined symmetry $\hat{M}_y \hat{T}$  for arbitrary $\theta$ with 
    \begin{equation}
    	U_{\hat{M}_y \hat{T}}=\begin{pmatrix}
    		0&-i\sigma_y&0\\
    		-i\sigma_y&0&0\\
    		0&0&-i\sigma_y
    	\end{pmatrix} i\sigma_y K=\begin{pmatrix}
    	0&\sigma_0&0\\
     \sigma_0&0&0\\
     0&0&\sigma_0
    	\end{pmatrix}K
     \end{equation}
    and
    \begin{equation}
    	U_{\hat{M}_y \hat{T}} H_{\text{SDW}}(-k_x,k_y)	U_{\hat{M}_y \hat{T}}^{-1}=H_{\text{SDW}}(k_x,k_y).
    \end{equation}

   \begin{comment}

    The full form of the Hamiltonian is 
    \begin{equation}
    	H_{\text{SDW}}(\bm{k})=\begin{pmatrix}
    		v_F(-\frac{\sqrt{3}}{2}k_x-\frac{1}{2}k_y)& \Delta(\frac{i}{2} \sigma_x-\frac{1}{2} \sigma_z) e^{-i\theta} &\Delta (\frac{i}{4}\sigma_x+\frac{\sqrt{3}}{4}i\sigma_y-\frac{1}{2}\sigma_z)e^{i\theta}\\
    		\Delta(-\frac{i}{2} \sigma_x-\frac{1}{2} \sigma_z) e^{i\theta} &v_F(\frac{\sqrt{3}}{2}k_x-\frac{1}{2}k_y) & \Delta (-\frac{i}{4}\sigma_x+\frac{\sqrt{3}}{4}i\sigma_y-\frac{1}{2}\sigma_z) e^{-i\theta} \\
    		\Delta (-\frac{i}{4}\sigma_x-\frac{\sqrt{3}}{4}i\sigma_y-\frac{1}{2}\sigma_z)e^{-i\theta}&\Delta(\frac{i}{4}\sigma_x-\frac{\sqrt{3}}{4}i\sigma_y-\frac{1}{2}\sigma_z) e^{i\theta} &v_F k_y	
    	\end{pmatrix}.
    \end{equation}
    \end{comment}
   \begin{comment}
     \begin{equation}
   	H'_{\text{SDW}}(\bm{k})=\begin{pmatrix}
   		v_F(\frac{\sqrt{3}}{2}k_x+\frac{1}{2}k_y)& \Delta(-\frac{i}{2} \sigma_x-\frac{1}{2} \sigma_z) e^{i\theta} &\Delta (-\frac{i}{4}\sigma_x-\frac{\sqrt{3}}{4}i\sigma_y-\frac{1}{2}\sigma_z)e^{-i\theta}\\
   		\Delta(\frac{i}{2} \sigma_x-\frac{1}{2} \sigma_z) e^{-i\theta} &v_F(-\frac{\sqrt{3}}{2}k_x+\frac{1}{2}k_y) & \Delta (\frac{i}{4}\sigma_x-\frac{\sqrt{3}}{4}i\sigma_y-\frac{1}{2}\sigma_z) e^{i\theta} \\
   		\Delta (\frac{i}{4}\sigma_x+\frac{\sqrt{3}}{4}i\sigma_y-\frac{1}{2}\sigma_z)e^{i\theta}&\Delta(-\frac{i}{4}\sigma_x+\frac{\sqrt{3}}{4}i\sigma_y-\frac{1}{2}\sigma_z) e^{-i\theta} &-v_F k_y	
   	\end{pmatrix}.
   \end{equation}
\end{comment}
It can also be shown that
 the system  would respect $C_{2z}$ symmetry at $\theta=\frac{n\pi}{3}$:
   \begin{equation}
	C_{2z} H_{\text{SDW}}(\bm{k},\theta=\frac{n\pi}{3})C_{2z}^{-1}=H'_{\text{SDW}}(-\bm{k},\theta=\frac{n\pi}{3}).
\end{equation}
    In other words, the point group of this triple-Q SDW is $C_6$ at $\theta=\frac{n\pi}{3}$ and reduces to $C_3$ at other angles.

Being similar to the CDW analysis, we can perform a unitary transform and rewrite the Hamiltonian in the basis $(\ket{E_{1,0}}\otimes\ket{\uparrow}, \ket{E_{1,1}}\otimes\ket{\downarrow}, \ket{E_{1,-1}}\otimes\ket{\uparrow},\ket{E_{1,0}}\otimes\ket{\downarrow}, \ket{E_{1,1}}\otimes\ket{\uparrow},\ket{E_{1,-1}}\otimes\ket{\downarrow})^{T}$, which carries the eigenvalues of $(e^{-i\frac{\pi}{3}},e^{-i\frac{\pi}{3}},e^{i\frac{\pi}{3}}, e^{i\frac{\pi}{3}}, 1,1)$ under $C_3$ operation. 
Then we obtain the effective Hamiltonian
   \begin{equation}
   	H^{\text{eff}}_{\text{SDW}}(\bm{k})=\begin{pmatrix}
  	H_{11} & -\frac{i}{2}k_{-}\sigma_z&-\frac{i}{2}k_{-}\sigma_z\\
   		\frac{i}{2}k_{+}\sigma_z&H_{22}&-\frac{i}{2}k_{-}\sigma_0\\
   		\frac{i}{2}k_{+}\sigma_z&\frac{i}{2}k_{+}\sigma_0&H_{33}
   	\end{pmatrix},
   \end{equation}
where the terms 
\begin{equation}
\begin{split}
    H_{11}(\theta)&=		-\frac{\sqrt{3}}{2}\Delta\cos(\theta+\frac{\pi}{6})\sigma_0+\Delta\cos(\theta+\frac{\pi}{6})\sigma_x\\
    &-\frac{\Delta}{2}\sin(\theta+\frac{\pi}{6})\sigma_z,\\ H_{22}(\theta)&=\frac{\sqrt{3}}{2}\Delta \cos(\theta-\frac{\pi}{6})\sigma_0+\Delta\cos(\theta-\frac{\pi}{6})\sigma_x\\
    &+\frac{\Delta}{2}\sin(\theta-\frac{\pi}{6})\sigma_z,\\H_{33}(\theta)&=-\frac{\sqrt{3}}{2}\Delta\sin\theta\sigma_0-\Delta\sin\theta\sigma_x+\frac{\Delta}{2}\cos\theta\sigma_z.
\end{split}
\end{equation}
 \section{Single-Q and double-Q states}

\begin{table*}[h]
  \caption{The comparison between density waves with different numbers of Q-vectors. }
    \centering
    \begin{tabular}{c|c|c|c}
    \hline\hline
   CDW &single-Q & double-Q& triple-Q \\\hline
     Oder parameter&$\Delta(\bm{r})=2\Delta\cos(Q_x x)$    & $\Delta(\bm{r})=2\Delta\sum_{\nu=x,y}\cos(\bm{Q_{\nu}}\cdot \bm{r})$&  $\Delta(\bm{r})=2\Delta\sum_{\nu=1}^3\cos(\bm{Q_{\nu}}\cdot \bm{r}+\theta)$\\
    hot spots&   \begin{minipage}{.2\textwidth}
      \includegraphics[width=0.6\linewidth]{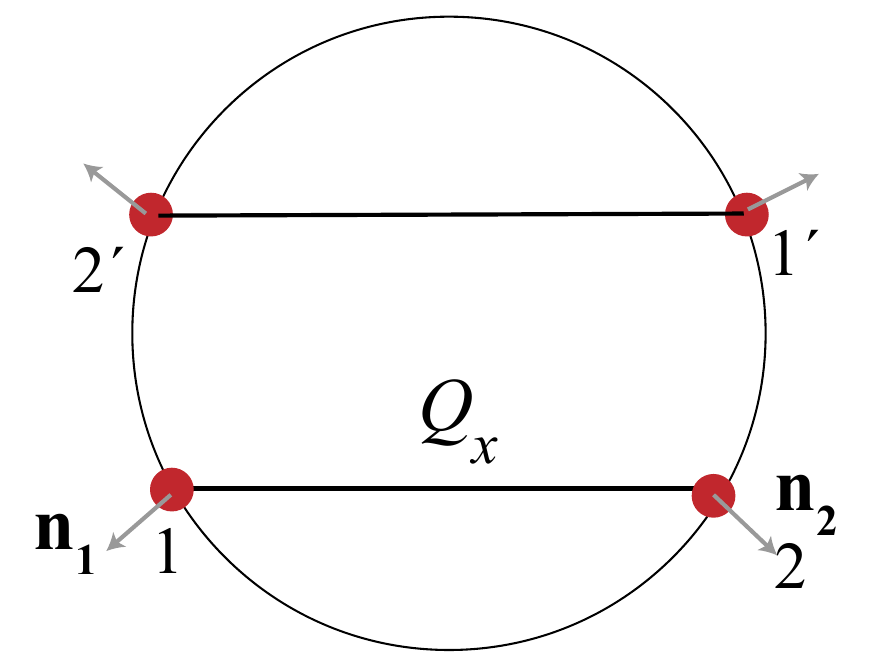}
    \end{minipage} & \begin{minipage}{.2\textwidth}
      \includegraphics[width=1\linewidth]{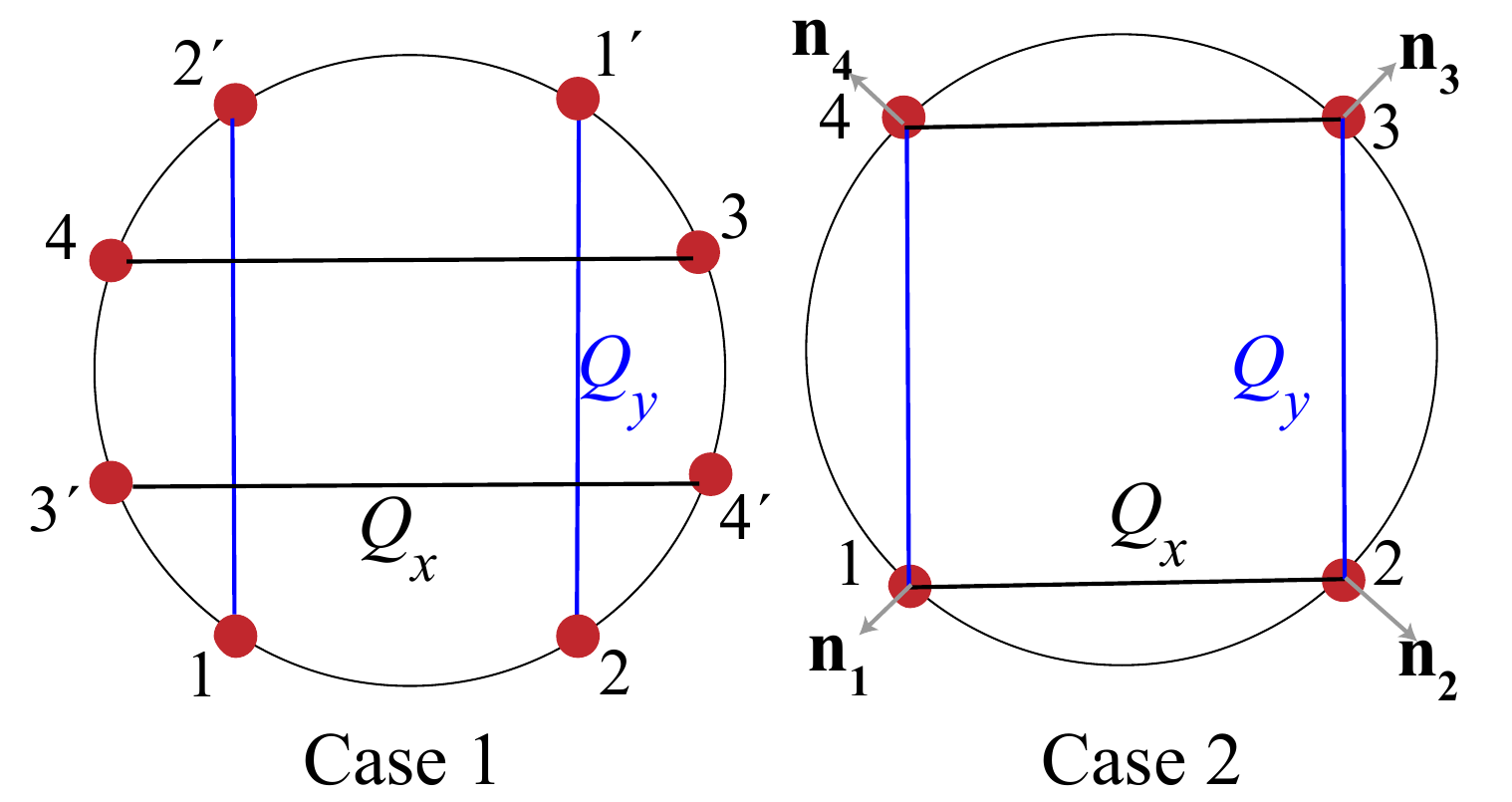}
    \end{minipage} &\begin{minipage}{.2\textwidth}
      \includegraphics[width=0.5\linewidth]{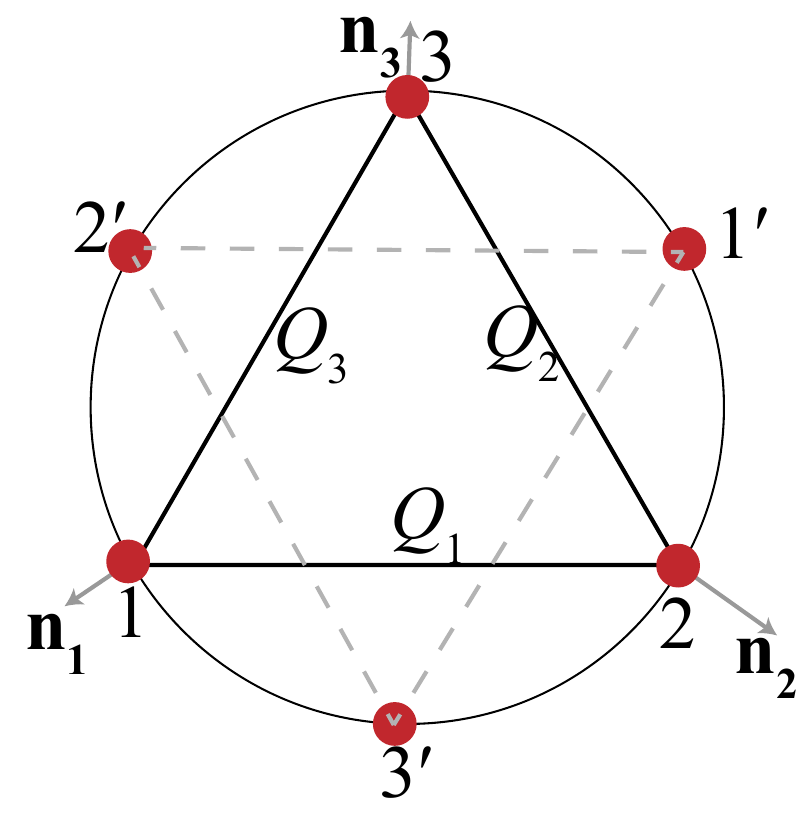}
    \end{minipage} \\
   low-energy Hamiltonian& non-Dirac& \makecell{case 1: non-Dirac  \\ case 2: multi-band Dirac} & multi-band Dirac\\\hline\hline
    \end{tabular}
  
    \label{tab2}
\end{table*}

After identifying the low-energy Dirac Hamiltonians induced by the triple-Q order parameters, one natural question is whether such low-energy Dirac Hamiltnmoians can appear in other density waves systems with different numbers of $Q$-vectors, such as single Q and double-Q vectors. As an exploration, we consider the single-Q CDW order parameter $\Delta(\bm{r})=2\Delta\cos(Q_x x)$ and double-Q CDW order parameter $\Delta(\bm{r})=2\Delta\sum_{\nu=x,y}\cos(\bm{Q}_{\nu}\cdot\bm{r})$. It is worth noting that, unlike the triple-Q density waves,  the single-Q and double-Q density waves do not have internal degrees of freedom.  While phasons can be excited in the moving density waves,  the phase shift $\theta$ is absent in the static single-Q and double-Q density waves.     
The hot spots connected by the single-Q and double-Q vectors on the Fermi circle are highlighted in Supplementary Table~\ref{tab2}. Following the previous section, we can perform a parallel analysis of the low-energy effective Hamiltonian of single/double-Q case. In the single-Q case, the  effective  Hamiltonian defined in the band basis at $\bm{k}=0$ is 
\begin{equation}
\mathcal{H}^{1Q}_{\text{eff}}(\bm{k})=\begin{pmatrix}
    -\Delta-k_y\sin\alpha &k_x\cos\alpha\\
    k_x\cos\alpha&\Delta-k_y\sin\alpha
\end{pmatrix}
\end{equation}
Here, the angle $\alpha$ labels the unit vector direction along the hot spot momenta. 
Evidently, it is non-Dirac.

In the double-Q case, there would be four hot spots on the Fermi circles connected by either $Q_x$ or $Q_y$ vectors (see Supplementary Table~\ref{tab2}).   We shall separate it into two cases:  (i) the hot spots on the Fermi surface connected by $Q_x$ and $Q_y$ are different; (ii) the hot spots on the Fermi surface connected by $Q_x$ and $Q_y$ are the same. Note that case 1 can be continuously tuned into case 2 by changing the parameters, such as the chemical potential. In the former case, every two hot spots give a 1D effective Hamiltonian as $\mathcal{H}^{1Q}_{\text{eff}}(\bm{k})$ so that the Dirac features are absent. In the second case, all four hot spots are connected by $Q$-vectors (see Supplementary Table~\ref{tab2}) and the effective Hamiltonian can display multi-band Dirac behaviors.

We now look at the second case in more detail.  To be specific, we set $Q_x=(1,0)Q, Q_y=(0,1)Q$. 
In the double-Q case (the case 2 in Table~\ref{tab2}),
\begin{equation}
    H_{\text{CDW}}^{2Q}(\bm{k})=\begin{pmatrix}
    v_F\bm{k}\cdot \hat{n}_1&\Delta &0&\Delta \\
    \Delta& v_F\bm{k}\cdot\hat{\bm{n}}_2&\Delta &0\\
    0&\Delta & v_F\bm{k}\cdot\bm{\hat{n}}_3&\Delta\\
    \Delta &0&\Delta & v_F\bm{k}\cdot\hat{\bm{n}}_4
    \end{pmatrix}.
\end{equation}
Here,  the unit vectors $\hat{\bm{n}}_1=\frac{1}{\sqrt{2}}(-1,-1), \hat{\bm{n}}_2=\frac{1}{\sqrt{2}}(1,-1), \hat{\bm{n}}_3=\frac{1}{\sqrt{2}}(1,1), \hat{\bm{n}}_4=\frac{1}{\sqrt{2}}(-1,1)$.
At $\bm{k}=0$, the eigenenergies are $E_{1}=2\Delta$, $E_{2}=0$,  $E_{3}=-2\Delta$, $E_{4}=0$. The corresponding eigenstates are 
\begin{equation}
    \begin{split}
      \ket{E_1}&=\frac{1}{2}(\ket{p_1}+\ket{p_2}+\ket{p_3}+\ket{p_4}),\\   \ket{E_2}&=\frac{1}{2}(i\ket{p_1}-\ket{p_2}-i\ket{p_3}+\ket{p_4}), \\\ket{E_3}&=\frac{1}{2}(-\ket{p_1}+\ket{p_2}-\ket{p_3}+\ket{p_4}), \\\ket{E_4}&=\frac{1}{2}(-i\ket{p_1}-\ket{p_2}+i\ket{p_3}+\ket{p_4}).  
    \end{split}
\end{equation}
 Projecting $H_{\text{CDW}}^{2Q}(\bm{k})$ into the basis spanned by ($\ket{E_1}, \ket{E_2}, \ket{E_3}, \ket{E_4}$), we obtain a four-band Dirac Hamiltonian:
\begin{eqnarray}
\mathcal{H}_{\text{eff}}^{2Q}(\bm{k})=\begin{pmatrix}
2\Delta& \tilde{v}_Fk_{+}&0& \tilde{v}^*_Fk_{-}\\
\tilde{v}^*_F k_{-}&0& \tilde{v}_F k_{+}&0\nonumber\\
0&\tilde{v}_F^* k_{-}&-2\Delta& \tilde{v}_F k_{+}\\
\tilde{v}_F k_{+}&0 &\tilde{v}^*_F k_{-}&0
\end{pmatrix}.\\
\end{eqnarray}
 Here, $\tilde{v}_F=\frac{v_F}{2} e^{-i\frac{3\pi}{4}}$. In other words, the double-Q density waves can also induce the multi-band Dirac Hamiltonian and exhibit nontrivial band geometry. Note that $\mathcal{H}_{\text{eff}}^{2Q}(\bm{k})$ is the total Hamiltnoian for the double-Q CDW already, and its dimension is smaller than  $\mathcal{H}^{3Q}_{\text{CDW},t}(\bm{k})$. It can be seen that $\mathcal{H}_{\text{eff}}^{2Q}(\bm{k})$ is a four-band Dirac Hamiltonian and exhibit nontrivial band geometry.

The comparison of density waves with different numbers of Q-vectors is summarized in Supplementary Table~\ref{tab2}. Our analysis shows that multiband Dirac Hamiltonians can be induced by multi-Q density wave order parameters as long as more than two hot spots are folded together simultaneously. This conclusion is expected to hold even when the CDW order parameter is replaced with SDW order parameters, although the effective dimension of the Hamiltonian would be doubled.

\section{Nonreciprocal transports in triple-Q SDWs}
\subsection{Magnetochiral anisotropy in  triple-Q SDWs}
 We consider a time-dependent drive
 \begin{equation}
 	\mathcal{E}(t)=\bm{E} \cos(\omega t).
 \end{equation} 
 Without loss of generality, we set the electric field $\bm{E}$ along $x$-direction. 
 In the experiment, it is convenient to identify the magnetochiral anisotropy by measuring the second harmonic generation of longitudinal conductivity \cite{Yokouchi2017}. Following the previous works \cite{Daniel2023,Binghai2024}, it can be deduced that the nonreciprocal longitudinal  charge transport can be given by 
\begin{equation}
\sigma_{\alpha\alpha\alpha}^{2\omega}=-\frac{e^3}{2\hbar^3(2i\omega+\tau^{-1})(i\omega+\tau^{-1})}\int \frac{d\bm{k}}{(2\pi)^d}\sum_n f_n^{(0)}\frac{\partial^3 \epsilon_n^{(0)}}{\partial k_\alpha^3}-\frac{2e^3 }{\hbar}\int \frac{d\bm{k}}{(2\pi)^d} f_n^{(0)}\partial_{k_\alpha} G_n^{\alpha\alpha}(2\omega)\label{Eq_S40}
\end{equation}
where $f_n ^{(0)}$ is the Fermi distribution for the n-th band $\epsilon^{(0)}_n$, and the band-renormalized quantum metric is defined as
\begin{equation}
	G_n^{ab}(2\omega)= -\frac{1}{4}\sum_{l\neq n} (\frac{\mathcal{A}_{nl}^a\mathcal{A}_{ln}^b}{-\hbar\omega+\epsilon_{ln}}+\frac{\mathcal{ A}^b_{nl} \mathcal{ A}^a_{ln}}{\hbar\omega+\epsilon_{ln}}).
\end{equation}
Here, $A^a_{mn}=\frac{\braket{n\bm{k}|\partial_a H|m\bm{k}}}{\epsilon_{nm}}$ denotes the Berry connection between $n-$th and $l-$th band with $\epsilon_{nm}=\epsilon^{(0)}_n-\epsilon^{(0)}_m$. In the DC transport limit, we are interested in $\omega \rightarrow 0$. In this case, the additional antiunitary symmetry $\hat{M}_{y}\hat{T}$   in triple-Q SDW would enforce $\sigma_{xxx}^{2\omega}=0$ by mapping $k_x$ to -$k_x$. Instead, we would focus on $\sigma_{yyy}^{2\omega}$ below. The results are summarized in Supplementary Fig.~\ref{fig:fig_S4}. It can be seen from (a) and (b) that $\sigma_{yyy}^{2\omega}$ is generally finite expect for at $\theta=\frac{n\pi}{3}$ due to the presence of  $C_{2z}$ symmetry. Note that we also find that the quantum metric contribution in Eq.~\ref{Eq_S40} is negligible.

 \begin{figure}[h]
	\centering
\includegraphics[width=0.8\linewidth]{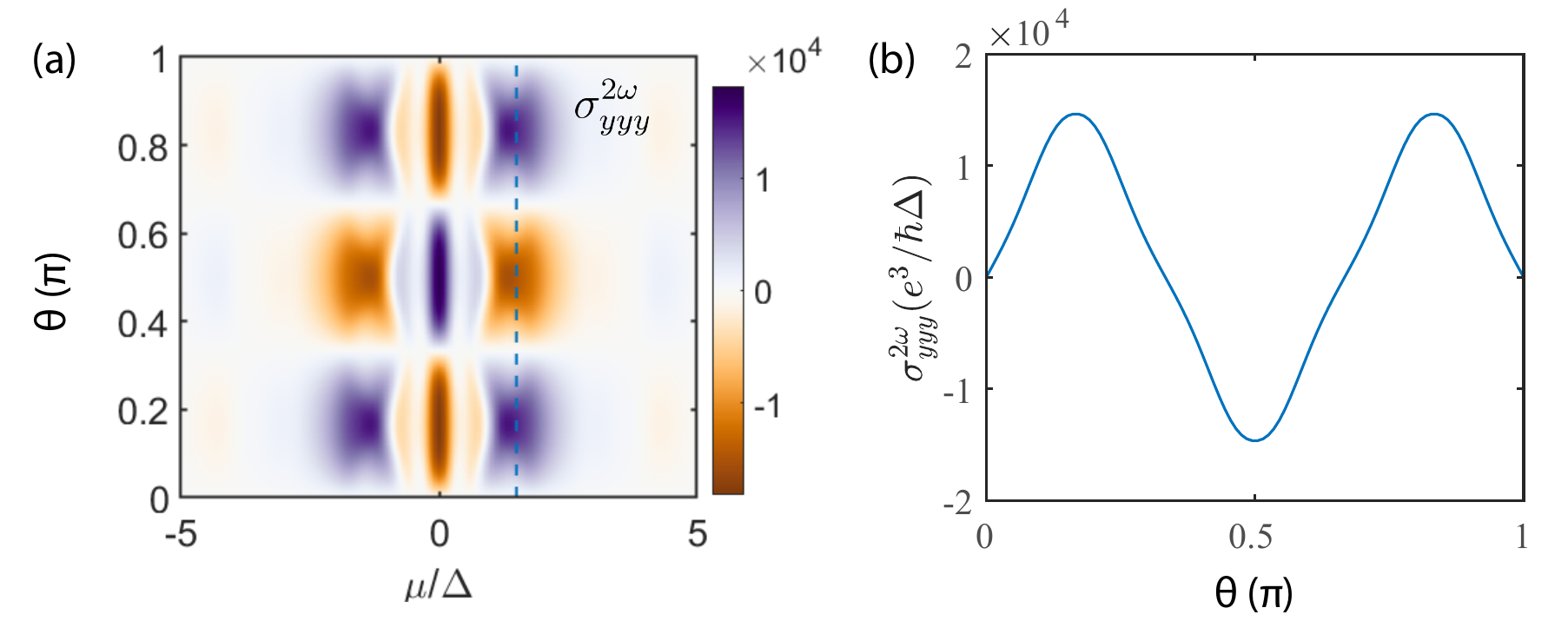}
	\caption{ The magnetochiral anisotropy represented by the strength of $\sigma^{2\omega}_{yyy}$ (in units of $\frac{e^3}{\hbar\Delta}$).(a) the phase shift dependence and chemical potential dependence of $\sigma_{yyy}^{2\omega}$ calculated   from the SDW effective Hamiltonian. (b) A line-cut of (a) at a fixed $\mu=1.5 \Delta$.  Here, the used parameters are $v_F=100$, $\Delta=1$, $T=0.01\Delta$, $\tau=0.01\Delta/\hbar$.}
	\label{fig:fig_S4}
\end{figure}

\subsection{Possible nonlinear Hall effects in strained triple-Q SDWs}
  \begin{table}[h]
  	\caption{Symmetry classification of momentum, spin, and strain tensor according to the irreducible representations of $C_3$ point group. }
  	\begin{tabular}{|c|c|c|}
  		\hline\hline
Irrep&Time-reversal Even&Time-reversal odd\\\hline
$A_1$& $u_{xx}+u_{yy}$& $\sigma_z$\\\hline
$E$& ($u_{xx}-u_{yy}$,$-2u_{xy}$)& $(k_x,k_y)$, $(\sigma_x,\sigma_y)$\\\hline
  	\end{tabular}
  \end{table}
 
In the main text, we mostly focus on the $C_3$ invariant case. However, the $C_3$ symmetry can be broken externally. The question is whether any interesting effects are allowed in this case. In this Supplementary Information section,   we demonstrate that the nonlinear Hall effects are possible in strained triple-Q  SDWs, such as strained SkX.

%  We consider the main effect of the  strain is to left energy degeneracy of three hot spots. The relevant degree of freedom in the three hot spots space can be captured by two Gell-Mann matrices
%  \begin{equation}
%  	\Gamma_0=\begin{pmatrix}
%  		1&0&0\\
%  		0&1&0\\
%  		0&0&1
%  	\end{pmatrix}, \Gamma_1=\begin{pmatrix}
%  		1&0&0\\
%  		0&-1&0\\
%  		0&0&0
%  	\end{pmatrix}, 	\Gamma_2=\frac{1}{\sqrt{3}}\begin{pmatrix}
%  	1&0&0\\
%  	0&1&0\\
%  	0&0&-2
%  	\end{pmatrix}.
%  \end{equation}
  
  To the leading order terms,  as shown in Table SI, the time-reversal invariant strain Hamiltonian is
  \begin{equation}
  	H_{strain}=v_F\sum_{\alpha\alpha} u_{\alpha} \Gamma_0\otimes \sigma_0+v_F[(u_{xx}-u_{yy})k_x-2u_{xy}k_y]\Gamma_0\otimes\sigma_z,
  \end{equation}
   where $\Gamma_0$ is a three-by-three identify matrix in the hot spots space,  the other group takes the same form. 
 The first term is to effectively shift the chemical potential so we can neglect it. The strain Hamiltonian is simplified as 
 \begin{equation}
 	H_{strain}\approx v_F[(u_{xx}-u_{yy})k_x-2u_{xy}k_y]\Gamma_0\otimes \sigma_z.
 \end{equation}
Without loss of generality, we can consider the uniaxial  strain tensor to be
 \begin{equation}
 	\begin{pmatrix}
 		u_{xx}&u_{xy}\\
 		u_{yx}&u_{yy}
 	\end{pmatrix}= u'\begin{pmatrix}
 	\cos^2\varphi-\nu \sin^2\varphi& (1+\nu) \cos\varphi\sin\varphi\\
 	(1+\nu)\cos\varphi\sin\varphi& -\nu\cos^2\varphi+\sin^2\varphi
 	\end{pmatrix}
 \end{equation}
 where $\varphi$ is to label the strain direction, $\nu$ is the Poisson ratio. 

 \begin{figure}
	\centering
\includegraphics[width=0.5\linewidth]{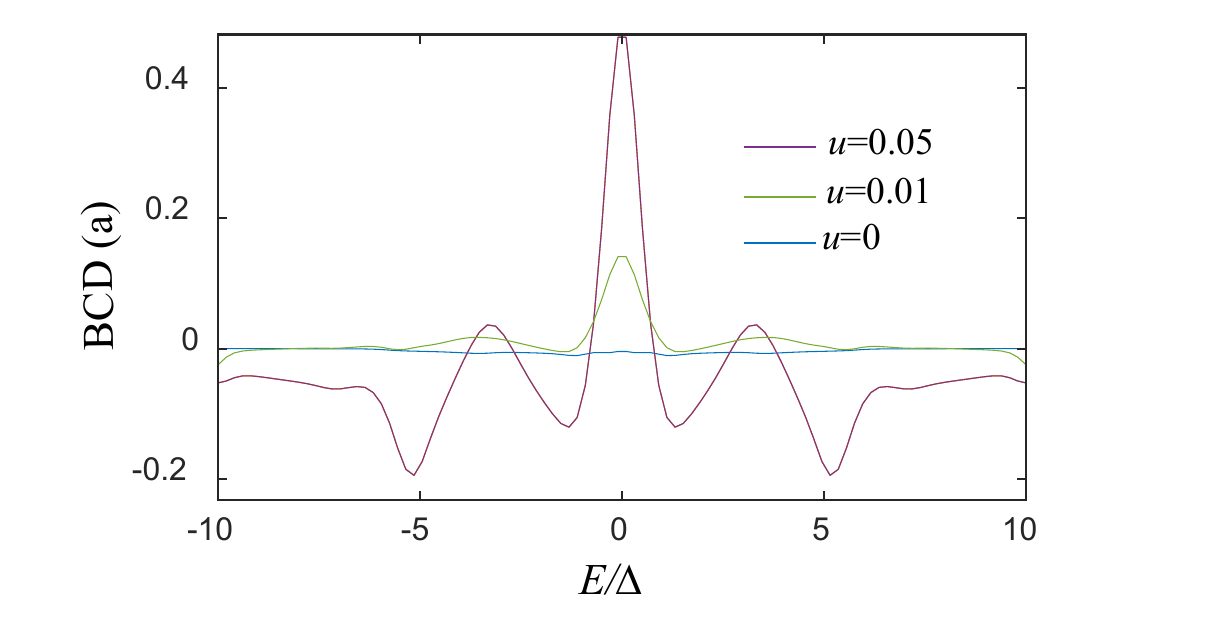}[h]
	\caption{Nonlinear Hall effects in strained SDW crystals. The Berry curvature dipole (BCD)  versus the energy $E$, where $E$ is artificially tuned from $[-10, 10]\Delta$ and $E=0$ is the Fermi energy. The strain strength is characterized by $u=0.05, 0.01, 0$, respectively. It can be seen that the BCD vanishes without strain due to the presence of $C_3$ symmetry, while there appears a sizable Berry curvature dipole when the SDW crystal is strained. Here, the used parameters in the triple-Q SDW effective Hamiltonian are $\Delta=1$, $v_F=100\Delta$, $\theta=0$, $\mu=0$. }
	\label{fig:fig_S2}
\end{figure}
 For simplicity, we set $\varphi=\pi/4$, and $u=(1+\nu)u'$, the strain Hamiltonian becomes
 \begin{equation}
	H_{strain}\approx -v_F u k_y \Gamma_0\otimes\sigma_z,
\end{equation}
 where $u$ characterizes the strain strength. Note that $u\neq 0$, the three-fold symmetry is broken. 
Then we can add the strain Hamiltonian into the total triple-Q Hamiltonian (note that we need to consider two sets of hot spots). The Berry curvature dipole is given by \cite{Liang2015}
  \begin{equation}
D_{\beta\gamma}=-\sum_{n} \int \frac{d\bm{k}}{(2\pi)^d}  \frac{\partial \epsilon_{n\bm{k}}}{\partial \bm{k}_{\beta}} \Omega_{n\gamma}(\bm{k}) \frac{\partial f(\epsilon_{n\bm{k}})}{\partial \epsilon_{n\bm{k}}}. \label{diople_BCD}
\end{equation}
The calculated strained induced finite Berry curvature dipole $D_{yz}$ is shown in Supplementary Fig.~\ref{fig:fig_S2}. It can be seen that there can support finite Berry curvature dipole in strained triple-Q SDW systems.

   \section{second harmonic generation  responses formalism}
\begin{figure}[h]
	\centering
	\includegraphics[width=0.9\linewidth]{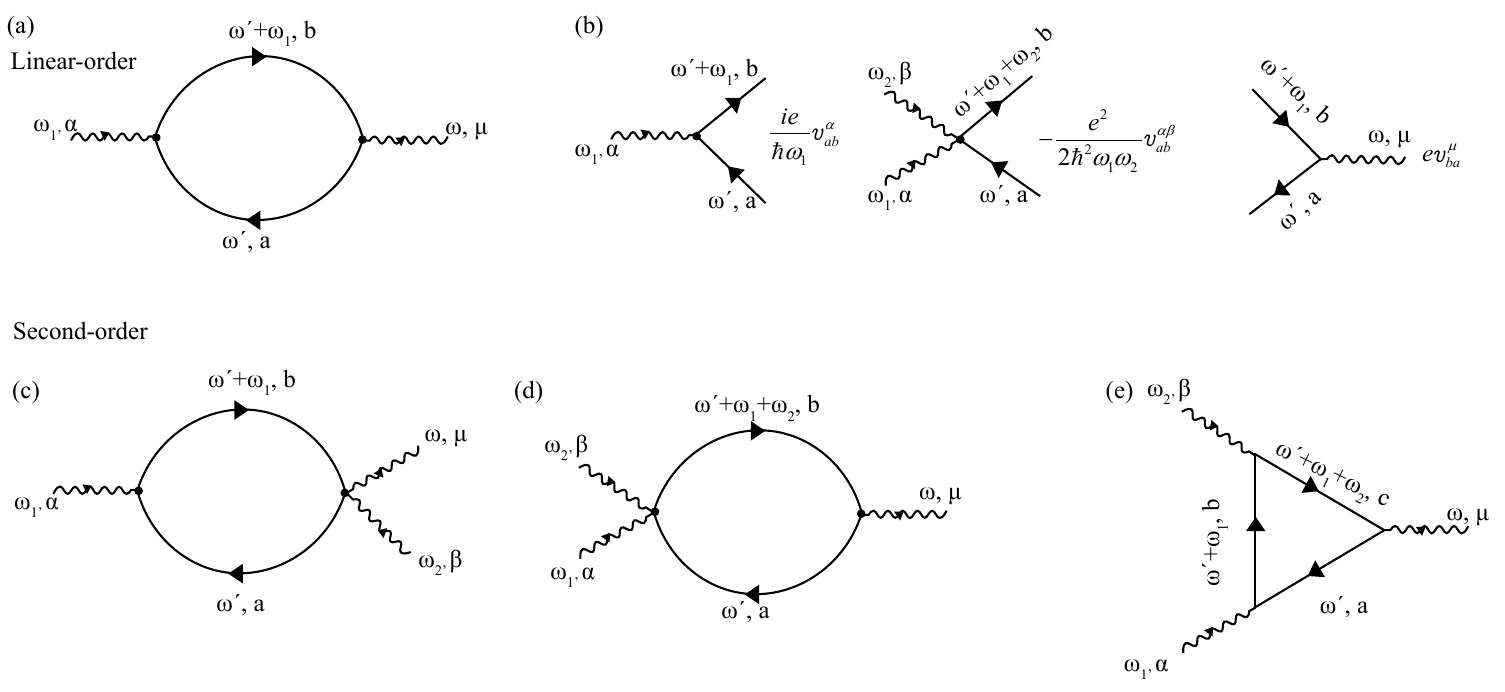}
	\caption{(a) The bubble diagram for the linear optical responses. (b) The vortex diagrams. (c) to (e) The Feynman diagrams for the second-order nonlinear optical response.}
	\label{fig:figs3}
\end{figure}
 We reproduce the optical responses formula given by ref.~\cite{Moore2019,  Takasan2021} here, which has been used in the main text to calculate the SHG below. The linear optical conductivity is given by the bubble diagram [Supplementary Fig.~3(a)],
\begin{eqnarray}
	\sigma^{\mu\alpha}(\omega;\omega)&&=\frac{ie^2}{\hbar\omega_1}\sum_{a\neq b}\int [d\bm{k}] \int d\omega' G_a(\omega') v_{ab}^{\alpha}G_{b}(\omega'+\omega_1)v_{ba}^{\mu}\nonumber\\
	&&=\frac{i e^2}{\hbar\omega}\sum_{a\neq b}\int [d\bm{k}] \frac{f_{ab} v_{ab}^{\alpha} v_{ba}^{\mu}}{\omega-\epsilon_{ba}}.
\end{eqnarray} 
where  $[d\bm{k}]=\int \frac{d\bm{k}}{(2\pi)^d}$, $v_{ab}^{\alpha}=\braket{a|\partial_{k_\alpha} H(\bm{k})|b}$, $H(\bm{k})$ is the Hamiltonian, the Fermi Dirac function difference between two bands $f_{ab}=f_a-f_b$, $\epsilon_{ba}=\epsilon_b-\epsilon_a$.

As shown in Supplementary Fig.~3, the second-order nonlinear optical conductivity is given by 
\begin{eqnarray}
	\sigma^{\mu\alpha\beta}(\omega_1+\omega_2;\omega_1,\omega_2)&&=-\frac{e^3}{2\hbar^2\omega_1\omega_2}\sum_{a,b,c}\int [d\bm{k}] \int d\omega' [G_a(\omega') v_{ab}^{\alpha} G_b(\omega'+\omega_1)v_{ba}^{\mu\beta}+\frac{1}{2}G_a(\omega')v_{ab}^{\alpha\beta} G_b(\omega'+\omega_1+\omega_2)v_{ba}^{\mu}+\nonumber\\
	&&G_a(\omega')v_{ab}^{\alpha}G_b(\omega'+\omega_1)v_{bc}^{\beta}G_c(\omega'+\omega_1+\omega_2)v^{\mu}_{ca}]+(\omega_1,\alpha)	\leftrightarrow (\omega_2,\beta),\nonumber\\
	&&=-\frac{e^3}{2\hbar^2\omega_1\omega_2}\int [d\bm{k}] [f_{ab}(\frac{v_{ab}^{\alpha}v_{ba}^{\mu\beta}}{\omega_1-\epsilon_{ba}}+\frac{v_{ab}^{\beta}v_{ba}^{\mu\alpha}}{\omega_2-\epsilon_{ba}})+f_{ab}\frac{v_{ab}^{\alpha\beta}v_{ba}^{\mu}}{\omega_{1}+\omega_{2}-\epsilon_{ab}}\nonumber\\
	&&+v_{ab}^{\alpha} v_{bc}^{\beta}
		 v_{ca}^{\mu}(\frac{f_{ab}}{(\omega_1-\epsilon_{ba})(\omega_1+\omega_2-\epsilon_{ca})}+\frac{f_{cb}}{(\omega_2-\epsilon_{cb})(\omega_1+\omega_2-\epsilon_{ca})})\nonumber\\
		 &&+v_{ab}^{\beta} v_{bc}^{\alpha}
		 v_{ca}^{\mu}(\frac{f_{ab}}{(\omega_2-\epsilon_{ba})(\omega_1+\omega_2-\epsilon_{ca})}+\frac{f_{cb}}{(\omega_1-\epsilon_{cb})(\omega_1+\omega_2-\epsilon_{ca})})].
\end{eqnarray}
Here, $v_{ab}^{\alpha\beta}=\braket{a|\partial_{k_\alpha}\partial_{k_\beta} H(\bm{k})|b}$.
For the second-harmonic generation ($\omega_1=\omega_2=\omega$),
\begin{eqnarray}
\sigma^{\mu\alpha\beta}(2\omega;\omega,\omega)=-\frac{e^3}{2\hbar^2\omega^2} \sum_{a,b,c}\int [d\bm{k}] [f_{ab}\frac{v_{ab}^{\alpha}v_{ba}^{\mu\beta}+v_{ab}^{\beta}v_{ba}^{\mu\alpha}}{\omega-\epsilon_{ba}}+f_{ab}\frac{v_{ab}^{\alpha\beta}v_{ba}^{\mu}}{2\omega-\epsilon_{ab}}\nonumber\\+(v_{ab}^{\alpha}v_{bc}^{\beta}+v_{ab}^\beta v_{bc}^{\alpha}) v_{ca}^{\mu}( \frac{f_{ab}}{(\omega-\epsilon_{ba})(2\omega-\epsilon_{ca})}+\frac{f_{cb}}{(\omega-\epsilon_{cb})(2\omega-\epsilon_{ca})})]
\end{eqnarray}
Using the identity:
\begin{equation}
 \frac{f_{ab}}{(\omega-\epsilon_{ba})(2\omega-\epsilon_{ca})}+\frac{f_{cb}}{(\omega-\epsilon_{cb})(2\omega-\epsilon_{ca})}=\frac{1}{\epsilon_{ab}+\epsilon_{cb}}[\frac{2f_{ac}}{2\omega-\epsilon_{ca}}+\frac{f_{cb}}{\omega-\epsilon_{cb}}+\frac{f_{ba}}{\omega-\epsilon_{ba}}].
\end{equation}

The optical conductivity for the second-harmonic generation can be rewritten as
\begin{eqnarray}
	\sigma^{\mu\alpha\beta}&&=\sigma_{\text{I}}^{\mu\alpha\beta}+\sigma_{\text{II}}^{\mu\alpha\beta},\\
	\sigma_\text{I}^{\mu\alpha\beta}&&=-\frac{e^3}{2\hbar^2\omega^2} \sum_{a\neq b}\int [d\bm{k}] f_{ab}\frac{v_{ab}^{\alpha}v_{ba}^{\mu\beta}+v_{ab}^{\beta}v_{ba}^{\mu\alpha}}{\omega-\epsilon_{ba}}+f_{ab}\frac{v_{ab}^{\alpha\beta}v_{ba}^{\mu}}{2\omega-\epsilon_{ab}}, \\
	\sigma_{\text{II}}^{\mu\alpha\beta}&&=-\frac{e^3}{2\hbar^2\omega^2} \sum_{a\neq b\neq c}\int [d\bm{k}] \frac{(v_{ab}^{\alpha}v_{bc}^{\beta}+v_{ab}^\beta v_{bc}^{\alpha}) v_{ca}^{\mu}}{\epsilon_{ab}+\epsilon_{cb}}(\frac{2f_{ac}}{2\omega-\epsilon_{ca}}+\frac{f_{cb}}{\omega-\epsilon_{cb}}+\frac{f_{ba}}{\omega-\epsilon_{ba}}),
\end{eqnarray}
where $2\omega$ and $\omega$ characterize the contributions from two-photon and one-photon processes, respectively.  The contribution that involves two (three) different bands is labeled as $\sigma_\text{I}^{\mu\alpha\beta}$ ($	\sigma_\text{II}^{\mu\alpha\beta}$), respectively.

 \end{document}